\documentclass[sigconf]{acmart}

\acmConference[SIGMOD '22]{SIGMOD '22: The 41st ACM SIGMOD International Conference on Management of Data}{June 12--17, 2022}{Philadelphia, PA}
\acmBooktitle{SIGMOD '22: The 41st ACM SIGMOD International Conference on Management of Data,
	June 12--17, 2022, Philadelphia, PA}
\acmPrice{15.00}
\acmISBN{978-1-4503-XXXX-X/18/06}

\copyrightyear{2022}
\acmYear{2022}
\setcopyright{acmcopyright}\acmConference[SIGMOD '22]{Proceedings of the 2022 International Conference on Management of Data}{June 12--17, 2022}{Philadelphia, PA, USA}
\acmBooktitle{Proceedings of the 2022 International Conference on Management of Data (SIGMOD '22), June 12--17, 2022, Philadelphia, PA, USA}
\acmPrice{15.00}
\acmDOI{10.1145/3514221.3517906}
\acmISBN{978-1-4503-9249-5/22/06}

\usepackage{graphicx}
\usepackage{tabularx}
\usepackage{url}
\usepackage{booktabs}
\usepackage{caption}
\usepackage{subcaption}
\usepackage{amsmath}

\usepackage[shortlabels]{enumitem}
\setlist[itemize]{leftmargin=*}
\setlist[enumerate]{leftmargin=*}

\usepackage[linesnumbered,ruled,vlined]{algorithm2e}

\newcommand{\arxiv}{1} % 0: SIGMOD Camera-ready 1: ArXiv
\usepackage{multirow}

\makeatletter
\def\BState{\State\hskip-\ALG@thistlm}
\makeatother

\newcommand{\name}{\mbox{\sc Doduo}}

\newcommand{\namest}{\mbox{\sc Dosolo}}
\newcommand{\namesc}{${\text{\mbox{\sc Dosolo}}}_{\text{\tiny {\sc SCol}}}$}

\newcommand{\hl}[1]{#1}
\newcommand{\hll}[1]{#1}
\newcommand{\hlll}[1]{#1}
\newcommand{\rev}[1]{#1}
\newcommand{\revcr}[1]{#1}
\begin{document}
\title{Annotating Columns with Pre-trained Language Models}

%%
%% The "author" command and its associated commands are used to define the authors and their affiliations.
% \settopmatter{authorsperrow=3}
% \author{}
% \affiliation{}
% \email{}

\settopmatter{authorsperrow=3}
\author{Yoshihiko Suhara, Jinfeng Li, Yuliang Li, Dan Zhang}
\affiliation{%
  \institution{Megagon Labs}
}
\email{{yoshi,jinfeng,yuliang,dan_z}@megagon.ai}

\author{\c{C}a\u{g}atay Demiralp}
\authornote{Work done while the author was at Megagon Labs.}
\affiliation{%
  \institution{Sigma Computing}
}
\email{cagatay@sigmacomputing.com}

\author{Chen Chen}\authornote{Deceased.}
\affiliation{%
  \institution{Megagon Labs}
}
\email{chen@megagon.ai}

\author{Wang-Chiew Tan}\authornotemark[1]
\affiliation{%
  \institution{Meta AI}
}
\email{wangchiew@fb.com}

\begin{abstract}
Inferring meta information about tables, such as column headers or relationships between columns, is an active
research topic in data management as we find many tables are missing some of this information.
In this paper, we study the problem of annotating table columns (i.e., predicting column types and the relationships between columns) using only information from the table itself. 
\hll{We develop a multi-task learning framework (called \name) based on pre-trained language models, which takes the entire table as input and predicts column types/relations using a single model.}
Experimental results show that \name{} establishes new state-of-the-art performance on two benchmarks for the column type prediction and column relation prediction tasks with up to 4.0\% and 11.9\% improvements, respectively. We report that \name{} can already outperform the previous state-of-the-art performance with a minimal number of tokens, only 8 tokens per column. 
\hll{We release a toolbox\footnote{\revcr{\url{https://github.com/megagonlabs/doduo}}}
and confirm the effectiveness of \name{} on a real-world data science problem through a case study.}
\end{abstract}

\maketitle

% %%% do not modify the following VLDB block %%
% %%% VLDB block start %%%
% \pagestyle{\vldbpagestyle}
% \begingroup\small\noindent\raggedright\textbf{PVLDB Reference Format:}\\
% \vldbauthors. \vldbtitle. PVLDB, \vldbvolume(\vldbissue): \vldbpages, \vldbyear.\\
% \href{https://doi.org/\vldbdoi}{doi:\vldbdoi}
% \endgroup
% \begingroup
% \renewcommand\thefootnote{}\footnote{\noindent
% This work is licensed under the Creative Commons BY-NC-ND 4.0 International License. Visit \url{https://creativecommons.org/licenses/by-nc-nd/4.0/} to view a copy of this license. For any use beyond those covered by this license, obtain permission by emailing \href{mailto:info@vldb.org}{info@vldb.org}. Copyright is held by the owner/author(s). Publication rights licensed to the VLDB Endowment. \\
% \raggedright Proceedings of the VLDB Endowment, Vol. \vldbvolume, No. \vldbissue\ %
% ISSN 2150-8097. \\
% \href{https://doi.org/\vldbdoi}{doi:\vldbdoi} \\
% }\addtocounter{footnote}{-1}\endgroup
% %%% VLDB block end %%%

% \ifdefempty{\vldbavailabilityurl}{}{
% \vspace{.3cm}
% \begingroup\small\noindent\raggedright\textbf{PVLDB Artifact Availability:}\\
% The source code, data, and/or other artifacts have been made available at \url{\vldbavailabilityurl} 
% %\jinfeng{comment out or add the source code link?}.
% \endgroup

\section{Introduction} \label{sec:intro}
%% Column annotation is important
\rev{Meta information about tables, such as column types and relationships between columns (or column relations), is crucial to a variety of data management tasks, 
including data quality control~\cite{Schelter:2018:DataQuality}, schema matching~\cite{Rahm:2001:SchemaMatching}, and data discovery~\cite{Chapman:2020:DatasetSearch}.
Recently, there is an increasing interest in identifying \emph{semantic} column
types and relations \cite{Hu:2019:CHI:VizNet,Hulsebos:2019:Sherlock,Zhang:2020:Sato}. 
Semantic column types such as 
``\textsf{population}'', ``\textsf{city}'', and ``\textsf{birth\_date}'' provide
contain finer-grained, richer information than standard DB types
such as integer or string.
Similarly, semantic column relations such as a binary relation 
``\textsf{is\_birthplace\_of}'' connecting a ``\textsf{name}'' and 
a ``\textsf{city}'' column can provide valuable information 
for understanding semantics of the table.
For example, commercial systems (e.g., Google Data Studio~\cite{GoogleDataStudio}
, Tableau~\cite{Tableau}) leverage such meta information for better table understanding. 
However, semantic column types and relations are typically missing in tables
while annotating such meta information manually can be quite expensive.
Thus, it is essential to build models that can automatically assign meta information to tables.}

\begin{figure}[t]
    \centering
    \includegraphics[width=0.45\textwidth]{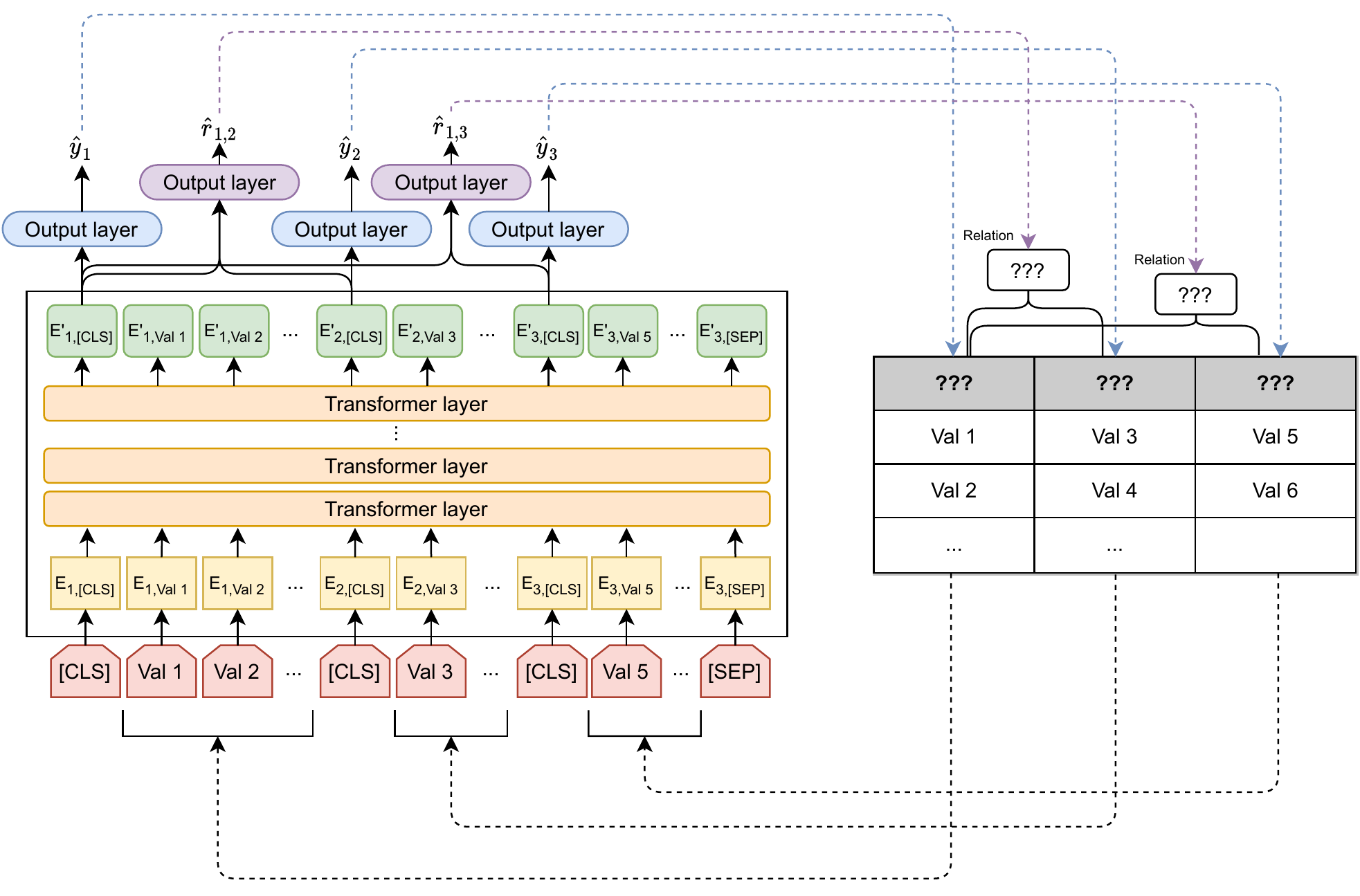}
    \caption{\rev{Overview of \name's model architecture.} \name{} serializes the entire table into a sequence of tokens to make it compatible with the Transformer-based architecture. 
    To handle the column type prediction and column relation extraction tasks, \name{} implements two different output layers on top of column representations and a pair of column representations, respectively.
    }\label{fig:model}
    \vspace{-1.5em}
\end{figure}

\begin{figure*}[th]
    \centering
    \includegraphics[width=0.6\textwidth]{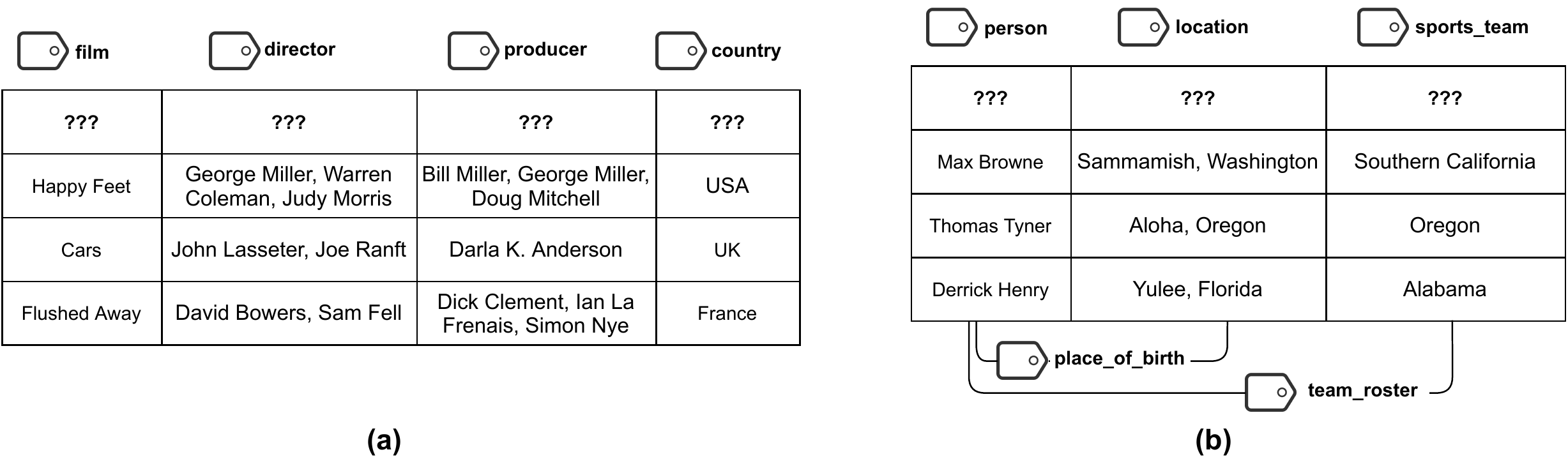}
    \caption{ Two example tables from the WikiTable dataset.
    (a) The task is to predict the column type of each column based on the table values.
    (b) The task is to predict both column types and relationships between columns.
    The column types (the column relations) are depicted at the top (at the bottom) of the table.
    This example also shows that column types and column relations are inter-dependent and hence, our motivation to develop a unified model for predicting both tasks.
    }\label{fig:examples}\vspace{-3mm}
\end{figure*}

%% Examples
%% Column type prediction
Figure~\ref{fig:examples} shows two tables with missing column types and column relations.
The table in Figure~\ref{fig:examples}(a) is about
animation films and the corresponding directors/producers/release countries of the films.  In the second and third columns, person names will require context, both in the same column and the other columns, to determine the correct column types. For example, {\tt George Miller}\footnote{In this context, {\tt George Miller} refers to an Australian filmmaker, but there exist more than 30 different Wikipedia articles that refer to different George Miller.} appears in both columns as a director and a producer, and it is also a common name. Observing other names in the column helps better understand the semantics of the column.
Furthermore, a column type is sometimes dependent on other columns of the table. Hence, by taking contextual information into account, the model can learn that the topic of the table is about (animation) films and understand that the second and third columns are less likely to be {\tt politician} or {\tt athlete}.
To sum up, this example shows that the table context and both intra-column and inter-column context can be very useful for column type prediction. 
%

%% Column relation prediction
Figure~\ref{fig:examples}(b) depicts a table with predicted column types and column relations. The column types {\tt person} and {\tt location} are helpful for predicting the relation {\tt place\_of\_birth}. However, 
it will still need further information to distinguish whether the location is {\tt place\_of\_birth} or {\tt place\_of\_death}.

The example above shows that column type and column relation prediction tasks are intrinsically related. Thus it will be synergistic to solve the two tasks simultaneously using a single framework.
%
%% Key challenges
% (1) column representation 
% (2) table context
% (3) how to combine column, column-pair features
To combine the synergies of column type prediction and column relation prediction tasks, we develop \name{} that:
(1) learns column representations, (2) incorporates table context, and (3) uniformly handles both column annotation tasks. Most importantly, our solution (4) shares knowledge between the two tasks. 

%% Model explanation
\name{} leverages a pre-trained Transformer-based language models (LMs) and adopts {\em multi-task learning} into the model to appropriately ``transfer'' shared knowledge from/to the column type/relation prediction task.
The use of the pre-trained Transformer-based LM makes \name{} \rev{a data-driven representation learning system}\footnote{In other words, \name{} relies on the general knowledge obtained from text corpora (e.g., Wikipedia) and a training set of tables annotated with column types and relations.} (i.e., feature engineering and/or external knowledge bases are not needed) (Challenge 1.)
%
% (2) table context
Pre-trained LM's contextualized representations and our table-wise serialization enable \name{} to naturally incorporate table context into the prediction (Challenge 2) and to handle different tasks using a single model (Challenge 3.)
%
% (3) easily incorporate column/column-pair features
Lastly, training such a {\em table-wise} model via multi-task learning helps ``transfer'' shared knowledge from/to different tasks (Challenge 4.)

Figure~\ref{fig:model} depicts the model architecture of \name. \name{} takes as input values from multiple columns of a table after serialization and predicts column types and column relations as output.
\name{} considers the table context by taking the serialized column values of all columns in the same table.
This way, both {\em intra-column} (i.e., co-occurrence of tokens within the same column) and {\em inter-column} (i.e., co-occurrence of tokens in different columns) information is accounted for.
\name{} appends a dummy symbol \texttt{[CLS]} at the beginning of each column and uses the corresponding embeddings as {\em learned column representations} for the column. The output layer on top of a column embedding (i.e., {\tt [CLS]}) is used for column type prediction, whereas the output layer for the column relation prediction takes the column embeddings of each column pair.

\smallskip
\noindent
{\bf Contributions~}
Our contributions are:
\begin{itemize}
    \setlength{\parskip}{0cm}
    \setlength{\itemsep}{0cm}
    \item We develop \name, a unified framework for both column type prediction and column relation prediction. \name{} incorporates table context through the Transformer architecture and is trained via multi-task learning. 
    \item Our experimental results show that \name\ establishes new state-of-the-art performance on two benchmarks, namely the WikiTable and VizNet datasets, with up to 4.0\% and 11.9\% improvements compared to TURL and Sato.
    \item We show that \name{} is data-efficient as it requires less training data or less input data. \name{} achieves competitive performance against previous state-of-the-art methods using less than half of the training data or only using 8 tokens per column as input. 
    \item \hll{We release the codebase and models as a toolbox, which can be used with just a few lines of Python code. We test the performance of the toolbox on a real-world data science problem and verify the effectiveness of \name{} even on out-domain data.}
\end{itemize}

\section{Related work}\label{sec:related}
%% Existing techniques
Existing column type prediction models enjoyed the recent advances in machine learning by formulating column type prediction as a multi-class classification task.
Hulsebos et al.~\cite{Hulsebos:2019:Sherlock} developed a deep learning model called Sherlock, which applies neural networks on multiple feature sets such as word embeddings, character embeddings, and global statistics extracted from individual column values.
Zhang et al.~\cite{Zhang:2020:Sato} developed Sato, which extends Sherlock by incorporating table context and structured output prediction to better model the nature of the correlation between columns in the same table.
Other models such as ColNet~\cite{Chen:2019:AAAI:ColNet}, HNN~\cite{Chen:2019:IJCAI:HNN}, Meimei~\cite{Takeoka:2019:AAAI:Meimei}, $C^2$~\cite{Khurana:2020:SemanticAnnotation4TabularData} use external Knowledge Bases (KBs) on top of machine learning models to improve column type prediction.
Those techniques have shown success on column type prediction tasks, outperforming classical machine learning models.

%% Column relation extraction
While those techniques identify the semantic types of individual columns, 
another line of work focuses on {\em column relations} between pairs of columns in the same table for better understanding tables~\cite{Limaye:2010:VLDB:AnnotatingWebTables,Venetis:2011:VLDB:RecoveringSemantics,Munoz:2014:WSDM:UsingLinkedData,Cannaviccio:2018:WWW:TowardsAnnotationgRelational,Macdonald:2020:CIKM:NeuralRelationExtraction,Deng:2020:TURL}.
A column relation is a semantic label between a pair of columns in a table, which offers more fine-grained information about the table. For example, a relation {\tt place\_of\_birth} can be assigned to a pair of columns {\tt person} and {\tt location} to describe the relationship between them.
Venetis et al.~\cite{Venetis:2011:VLDB:RecoveringSemantics} use an Open IE tool~\cite{Yates:2007:NAACL:TextRunner} to extract triples to find relations between entities in the target columns. 
Mu\~{n}oz et al.~\cite{Munoz:2014:WSDM:UsingLinkedData} use machine learning models to filter triple candidates created from DBPedia. 
Cannaviccio et al.~\cite{Cannaviccio:2018:WWW:TowardsAnnotationgRelational} use a language model-based ranking method~\cite{Zhai:2008:LM4IR}, which is trained on a large-scale web corpus, to re-rank relations extracted by an open relation extraction tool~\cite{Nakashole:2012:PATTY}.
Cappuzo et al.~\cite{Cappuzzo:2020:SIGMOD:CreatingEmbeddings} represent table structure as a graph and then learn the embeddings from the descriptive summaries generated from the graph.

%% Pre-trained Transformers for Data Management tasks
Recently, pre-trained  Transformer-based Language Models (LMs) such as BERT, which were originally designed for NLP tasks, have shown success in data management tasks. Li et al.~\cite{Li:2020:Ditto} show that pre-trained LMs is a powerful base model for entity matching. Macdonald et al.~\cite{Macdonald:2020:CIKM:NeuralRelationExtraction} proposed applications for entity relation detection. Tang et al.~\cite{Tang:2020:RelationalPretrained} propose RPTs as a general framework for automating human-easy data preparation tasks like data cleaning, entity resolution and information extraction using pre-trained masked language models.
The power of Transformer-based pre-trained LMs can be summarized into two folds. First, using a stack of Transformer blocks (i.e., self-attention layers), the model is able to generate contextualized embeddings for structured data components like table cells, columns, or rows. Second, models pre-trained on large-scale textual corpora can store ``semantic knowledge'' from the training text in the form of model parameters. For example, BERT might know that George Miller is a director/producer since the name frequently appears together with ``directed/produced by'' in the text corpus used for pre-training. 
In fact, recent studies have shown that pre-trained LMs store a significant amount of factual knowledge, which can be retrieved by template-based queries~\cite{Petroni:2019:LAMA,Jiang:2020:HowCanWeKnowLM,Roberts:2020:HowMuchKnowledge}.
%

%% For tables
Those pre-trained models have also shown success in data management tasks on tables.
TURL~\cite{Deng:2020:TURL} is a Transformer-based pre-training framework for table understanding tasks. Contextualized representations for tables are learned in an unsupervised way during pre-training and later applied to 6 different tasks in the fine-tuning phase. 
SeLaB~\cite{Trabelsi:2020:SeLaB} leverages pre-trained LMs for column annotation while incorporating table context. Their approach uses fine-tuned BERT models in a two-stage manner.
TaPaS\cite{herzig2020tapas} conducts weakly supervised parsing via pre-training, and TaBERT\cite{yin2020tabert} pre-trains for a joint understanding of textual and tabular data \rev{for the text-to-SQL task.}
TUTA~\cite{Wang2:020:StructureAwarePretraining4Table} makes use of different pre-training objectives to obtain representations at token, cell, and table levels and propose a tree-based structure to describe spatial and hierarchical information in tables. 
TCN~\cite{Wang:2021:TCN} makes use of both information within the table 
and \rev{across external tables from similar domains} to predict column type and pairwise column relations.

\rev{
In this paper, we empirically compare \name{} with Sherlock~\cite{Hulsebos:2019:Sherlock}, Sato~\cite{Zhang:2020:Sato}, and TURL~\cite{Deng:2020:TURL} as baseline methods. 
Sherlock is a single-column model while \name{} is multi-column by
leveraging table context to predict column
types and relations more accurately. 
Sato leverages topic model (LDA) features
as table context while \name{} can additionally take into account
fine-grained, token-level interactions among columns via its built-in self-attention 
mechanism. 
TURL is also a Transformer-based model like \name{} but it requires
additional meta table information such as table headers for pre-training. \name{} is more generic as it predicts column types and relations only relying on cell values in the table. See Section \ref{sec:experiment} for a more detailed comparison. }

\section{Background}\label{sec:background}
In this section, we formally define the two column annotation tasks: column type prediction and column relation annotation.
We also provide a brief background on pre-trained language models (LMs) and how to fine-tune them for performing column annotations.

\begin{table}[t]
    \centering
    \caption{Notations.}\vspace{-1em}\label{tab:notation}
    {\footnotesize
    \begin{tabular}{c|c}\hline
      Symbol & Description \\\hline\hline
      $T = (c_1, c_2, \dots, c_{n})$ & Columns in a table. \\\hline
      $c_i = (v^i_1, v^i_2, \dots, v^i_{m})$ & Column values. \\\hline
      $v^i_j = (w^i_{j,1}, w^i_{j,2}, \dots, w^i_{j, K} )$ & A single column value. \\\hline
      $D_{\rm train} = \left\{ T^{(n)}, L_{\rm type}^{(n)}, L_{\rm rel}^{(n)} \right\}_{n=1}^N$ & Training data \\\hline
      $L_{\rm type} = (l_1, l_2, \dots, l_{n})$, 
      $l_* \in \mathcal{C}_{\rm type}$ & Column type labels. \\\hline
      $L_{\rm rel} = (l_{1,2}, l_{1,3}, \dots, l_{1, {n}})$, 
      $l_{*,*} \in \mathcal{C}_{\rm rel}$ & Column relation labels.\\\hline
    \end{tabular}
    }
    \vspace{-1em}
\end{table}

\subsection{Problem Formulation}
%% Task definition
The goal of the column type prediction task is to classify each column to its \emph{semantic type}, such as
``country name'', ``population'', and ``birthday'' instead of the standard column types such as 
\texttt{string}, \texttt{int}, or \texttt{Datetime}. See also Figure~\ref{fig:examples} for more examples.
For column relation annotation, our goal is to classify the relation of each pair of columns.
In Figure \ref{fig:examples}, the relation between the ``person'' column and the ``location'' column
can be ``\texttt{place\_of\_birth}''.

\newcommand{\val}{\mathsf{val}}
\newcommand{\bfT}{\mathbf{T}}
\newcommand{\calC}{\mathcal{C}}
\newtheorem{problem}{Problem}

\rev{As summarized in Table~\ref{tab:notation}, more formally,} we consider a standard relational data model where a relation $T$ (i.e., table) consists of a set of attributes $T = (c_1, \dots c_n)$ (i.e., columns.) We denote by 
$\val(T.c_i)$ the sequence of data values stored at the column $c_i$. \rev{
For each value $v \in \val(T.c_i)$,
we assume $v$ to be of the \texttt{string} type and 
can be split
into a sequence of input tokens 
$v = [w_1, \dots, w_k]$ to pre-trained LMs.}
\rev{This approach of casting cell values into text
might seem restricted since tables
columns can be of numeric types such as \texttt{float} or 
\texttt{date}. There has been extensions of the Transformer
models to support numeric data~\cite{wu2020deep} 
and providing such direct support of numeric data is important
future work.
We also provide a brief analysis on \name{}'s performance
on numeric column types in Section \ref{sec:mainresult}.
}

\rev{
\begin{problem}[Column type prediction]
Given a table $T$ and a column $c_i$ in $T$,
a column type prediction model $\mathcal{M}$ 
with type vocabulary $\calC_{\rm{type}}$
predicts a column type
$\mathcal{M}(T, c_i) \in \calC_{\rm{type}}$
that best describes the semantics of $c_i$.
\end{problem}

{
\begin{problem}[Column relation prediction]
Given a table $T$ and a pair of columns
$(c_i, c_j)$ in $T$, 
a column relation prediction model $\mathcal{M}$
with relation vocabulary $\calC_{\rm{rel}}$
predicts a relation
$\mathcal{M}(T, c_i, c_j) \in \calC_{\rm{rel}}$ that best describes the semantics of the relation between $c_i$ and $c_j$.
\end{problem}
}
}

\rev{
In \name{}, we consider the supervised setting of 
multi-class classification. This means that we assume
a training data set $D_{\rm{train}}$ of tables 
annotated with columns types and relations from two fixed vocabularies
$(\calC_{\rm{type}}, \calC_{\rm{rel}})$.
Note that \name{} does not restrict itself to specific choices of vocabularies 
$(\calC_{\rm{type}}, \calC_{\rm{rel}})$ which are customizable by
switching the training set $D_{\rm{train}}$.
In practice, the choice of $(\calC_{\rm{type}}, \calC_{\rm{rel}})$ 
is ideally application-dependent. For example, if the downstream task
requires integration with a Knowledge Base (KB), it is ideal to have 
$(\calC_{\rm{type}}, \calC_{\rm{rel}})$ aligned with the KB's type/relation vocabulary.
In our experiment, we evaluated \name{} on datasets annotated with 
(1) KB types~\cite{bollacker2008freebase} and (2) DBPedia types~\cite{Nguyen:2020:MTab4DBPedia}.

The size and quality of the training set are 
also important for training high-quality 
column annotation models. While manually 
creating such datasets can be quite expensive, 
the datasets used in our experiments rely on 
heuristics that map table meta-data 
(e.g., header names, entity links)
to type names to create training sets
of large scale. See Section \ref{sec:datasets}
for more details.

While KB can work as a training example provider, \name{} does not require the training examples to be from a single source but can combine labels from any resources such as human annotations, labeling rules, and meta-data that can be transformed into the column type/relation label format.

We also note that the learning goal of \name{} is to train
column annotation models with high accuracy while being generalizable to unannotated tables (e.g., as measured by an unseen test set $D_{\rm{test}}$).
The column type/relation prediction models of \name{}
\emph{only} considers the table content (i.e., cell values) as input. 
This setting allows \name{} to be more flexible to practical applications
without replying on auxiliary information such as column names, table titles/captions, 
or adjacent tables typically required by existing works 
(See Section \ref{sec:related} for a comprehensive overview).
}

\begin{figure}[t]
    \centering
    \includegraphics[width=0.4\textwidth]{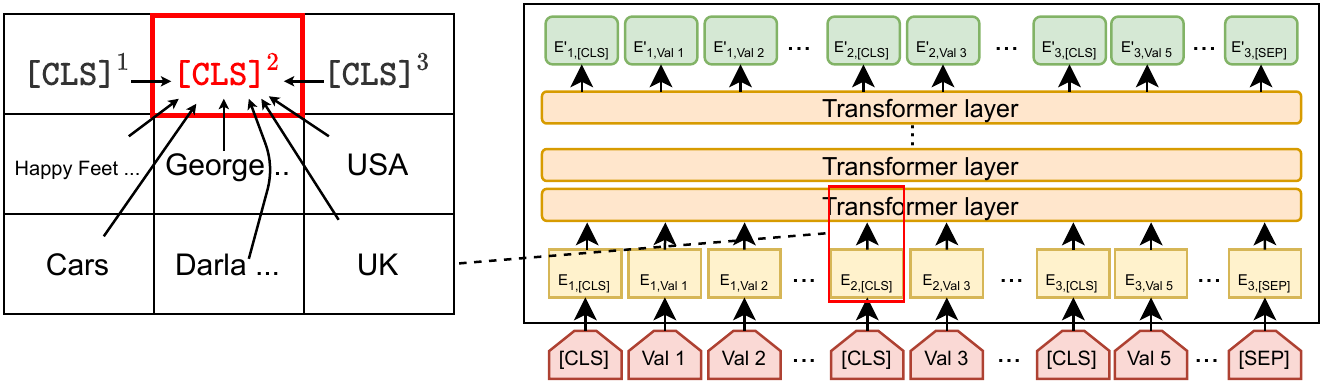}
    \caption{ How \name{} computes contextualized column embeddings using the Transformer layers. Each Transformer block calculates an embedding vector for every token based on surrounding tokens.}
    \label{fig:context_colemb}
    \vspace{-1.5em}
\end{figure}

\subsection{Pre-trained Language Models}
Pre-trained Language Models (LMs) emerge as general-purpose solutions to tackle various natural language processing (NLP) tasks. Representative LMs such as BERT~\cite{Devlin:2019:BERT} and ERNIE~\cite{SunWLFTWW20aaai} have shown leading performance among all solutions in NLP benchmarks such as GLUE~\cite{glue,WangSMHLB19}.
These models are pre-trained on large text corpora such as Wikipedia pages and typically employ multi-layer Transformer blocks~\cite{vaswani2017attention} to assign more weights to informative words and less weight to stop words for processing raw texts.  
During pre-training, a model is trained on self-supervised language prediction tasks such as missing token prediction and next-sentence prediction. The purpose is to learn the semantic correlation of word tokens (e.g., synonyms), such that correlated tokens can be projected to similar vector representations.  
After pre-training, the model is able to learn the lexical meaning of the input sequence in the shallow layers and the syntactic and semantic meanings in the deeper layers~\cite{Clark:2019:WhatDoesBERTLookAt,Tenney:2019:BertRediscovers}.

A special component of pre-trained LMs is the attention mechanism, which embeds a word into a vector based on its context (i.e., surrounding words). The same word has different vectors if it appears in different sentences, and this is very different from other embedding mechanisms such as word2vec~\cite{mikolov2013word2vec}, GloVe~\cite{pennington-etal-2014-glove}, and fastText~\cite{bojanowski2016enriching}, which always generate the same vector for the same word in any context. Pre-trained LMs' embeddings are context-dependent and thus offer two strengths. First, it can discern polysemy. For example, the person name {\tt George Miller} referring to a producer is different from the same name that refers to a director. Pre-trained LMs discern the difference and generate different vectors. Second, the embedding deals with synonyms well. For example, the words {\tt Derrick Henry} and {\tt Derrick Lamar Henry Jr} (respectively, ({\tt USA}, {\tt US}), ({\tt Oregon}, {\tt OR})) are likely the same given their respective contexts. Pre-trained LMs will generate similar word vectors accordingly. Due to the two favorable strengths, pre-trained models should enable the best performance to column annotation tasks, where each cell value is succinct, and its meaning highly depends on its surrounding cells.

The pre-trained model does not know what to predict for specific tasks unless the task is exactly the same the pre-training task. 
Thus, a pre-trained LM needs to be {\em fine-tuned} with task-specific training data, so the model can be tailored for the task. 
A task-specific output layer is attached to the final layer of the pre-trained LM, and the loss value (e.g., cross-entropy loss) is back-propagated from the output layer to the pre-trained LM for a minor adjustment.

In this paper, we fine-tune the popular 12-layer BERT Base model~\cite{Devlin:2019:BERT}. However, \name{} is independent of the choice of pre-trained LMs, and \name\ can potentially perform even better with larger pre-trained LMs.

\subsection{Multi-task learning}
Multi-task learning~\cite{Caruana:2004:MTL} is a type of supervised machine learning framework, where the objective function is calculated based on more than one task. Generally, different types of labels are used, and the labels may be or may not be annotated on the same example. 
The intuition and an assumption behind multi-task learning is that the tasks intrinsically share knowledge, and thus, training with the same base model benefits each other.

The major benefit of multi-task learning is that it can help improve the generalization performance of the model, especially when the training data is not sufficient. 
Multi-task learning can be easily applied to Deep Learning models~\cite{Ruder:2017:MTLOverview} by attaching different output layers to the main model, which is considered a ``learned'' representation encoder that converts input data to dense representations.

There are a variety of approaches for multi-task learning~\cite{Ruder:2017:MTLOverview}, depending on how to model and optimize shared parameters. 
Multi-task learning models can be split into two categories based on how parameters are shared. With {\em hard parameter sharing}~\cite{Caruana:1993:MTLHardParameterSharing}, models for multiple tasks share the same parameters, whereas {\em soft parameter sharing}~\cite{Yang:2017:ICLR:MTLSoftParameterSharing} adds constraints on distinct models for different tasks. 
In this paper, we consider hard parameter sharing as it is a more cost-effective approach. Among hard parameter sharing models, we choose a joint multi-task learning framework~\cite{Hashimoto:2017:EMNLP:JointMTL} that uses the same base model with different output layers for different tasks.

\section{Model} \label{sec:model}
In this section, we first introduce a baseline single-column model that fine-tunes a pre-trained LM on individual columns. Then, we describe \name{}'s model architecture and training procedure.

\subsection{Single-column Model}
Since LMs take token sequences (i.e., text)
as input, one first has to convert a table into token sequences so that they can be meaningfully processed by LMs.
A straightforward serialization strategy is to simply concatenate column values to make a sequence of tokens and feed that sequence as input to the model.
That is, suppose a column $C$ has column values $v_1, \dots v_m$, the serialized sequence is
{
\setlength\abovedisplayskip{2pt}
\setlength\belowdisplayskip{2pt}
$$ {\rm serialize_{\rm single}}(C) ::= \mathsf{[CLS]} \ v_1 \  \dots \ v_m  \mathsf{ [SEP] },$$
}
where $\mathsf{[CLS]}$ and $\mathsf{[SEP]}$ are special tokens used to mark the beginning and end of a sequence\footnote{Note that {\tt [CLS]} and {\tt [SEP]} are the special tokens for BERT and other LMs may have other special tokens, which are usually implemented as part of their tokenizers.}. 
For example, the first column of the first table in Figure~\ref{fig:examples} is serialized as: \textsf{[CLS] Happy Feet Cars Flushed Away [SEP]}.
This serialization converts the problem into a sequence classification task. Thus, it is straightforward to fine-tune a BERT model using training data.

% Column relation prediction
The column relation prediction task can be also formulated as a sequence classification task by converting a pair of columns (instead of a single column) into a token sequence in a similar manner.
For this case, we also insert additional $\mathsf{[SEP]}$ between values of two columns to help the pre-trained LM distinguish the two columns.
Namely, given two columns $C = v_1, \dots, v_m$ and $C' = v_1', \dots, v_m'$, the single-column model serializes the pair as:
{
\setlength\abovedisplayskip{2pt}
\setlength\belowdisplayskip{2pt}
$$ {\rm serialize_{\rm single}}(C, C') ::= \mathsf{[CLS]} \ v_1 \  \dots \ v_m  \mathsf{ [SEP] } \ v'_1 \  \dots \ v'_m  \mathsf{ [SEP] } .$$
}

\vspace{-1em}
Using the above serialization scheme, we can cast the column type and relation prediction tasks as sequence classification
and sequence-pair classification tasks, which can be solved by LM fine-tuning. However, such sequence classifications predict column types independently, even if they are in the same table. 
We refer to this method as the \emph{single-column} model.
Although the single-column model can leverage the language understanding capability and knowledge learned by the LM via pre-training, 
it has an obvious drawback of treating columns in the same table as independent sequences.
As a result, the single-column model fails to capture the table context, which is known to be important for the column annotation tasks~\cite{Chen:2019:IJCAI:HNN,Zhang:2020:Sato,Khurana:2020:SemanticAnnotation4TabularData}.

\subsection{Table Serialization}
%% Table serialization
In contrast to the single-column model described above, \name{} is a {\em multi-column} (or table-wise) model that takes an entire table as input.
\name\ serializes data entries as follows:
for each table that has $n$ columns $T = ( c_i )_{i=1}^{n}$, where each column has $N_m$ column values $c_i = (v^i_j)_{j=1}^{m}$. We let
{
\setlength\abovedisplayskip{2pt}
\setlength\belowdisplayskip{2pt}
$$ {\rm serialize}(T) ::= \mathsf{[CLS] } \ v^1_1 \  \dots \mathsf{ [CLS] } \ v^{n}_1 \ \dots v^{n}_{m} \mathsf{ [SEP] }.$$
}
For example, the first table in Figure~\ref{fig:examples} is serialized as:
\smallskip
\noindent
\begin{small}
\begin{center}
\textsf{[CLS] Happy Feet, \dots  [CLS] George Miller, \dots [CLS] USA, \dots, France [SEP]}.
\end{center}
\end{small}

Different from the single-column model, which always has a single $\mathsf{[CLS]}$ token in the input, \name's serialization method inserts as many $\mathsf{[CLS]}$ tokens as the number of columns in the input table. 
This difference makes a change in the classification formulation. While the single-column model classifies a single sequence (i.e., a single column) by predicting a single label, \name{} predicts as many labels as the number of $\mathsf{[CLS]}$ tokens in the input sequence.
\rev{The learning procedure of \name{} starts 
by serializing and tokenizing all tables in the datasets (Line 3-4 of Algorithm \ref{alg:training}.)}

\SetKwInOut{Parameter}{Variables}
\begin{algorithm}[!ht]
%\small
\footnotesize
\caption{Training procedure of \name}\label{alg:training}
\rev{
	\KwIn{Number of training epochs $N_{\rm Epoch}$; 
	For each task $i \in [1, T]$, training set $D_i$,
	loss function $\mathcal{L}_i$, and optimizer $O_i$}
    \KwOut{Annotation model $\mathcal{M} = (\mathsf{LM}, \{g_1, \dots, g_T\})$ with a language model $\mathsf{LM}$ and $T$ heads
    $g_1$ to $g_T$}
    
    \tcp{Initialize model weights}
    Initialize $\mathsf{LM}$ using its pre-trained weights\;
    Initialize $\{g_1, \dots, g_T\}$ randomly\;
    
    \tcp{Serialize all tables in each training set}
    \For{$i = 1$ to $T$} {
        $D_i \leftarrow \{\text{serialize}(T) \text{ for } T \in D_i\}$\;
    }
    
	\For{ $\mathsf{ep} = 1$ to $N_{\rm Epoch}$}{
	    \For{$i = 1$ to $T$} {
	        Randomly split $D_i$ into mini-batches $\{B_1, \dots B_k\}$\;
	        \For{$B$ in $\{B_1, \dots B_k\}$} {
	            \tcp{Evaluate the $\mathcal{L}_i$ loss func on the $g_i$ head}
	            $L \leftarrow \mathcal{L}_i(B, \mathsf{LM} \oplus g_i)$\;
                \tcp{Back-propagate to update model weights}
	            $\mathcal{M} \leftarrow \textsf{back-propagate}(\mathcal{M}, O_i, \partial L / \partial (\mathsf{LM} \oplus g_i)) $\;
	        }
	    }
	}
    \Return $\mathsf{M}$\;}
\end{algorithm}

\subsection{Contextualized Column Representations}
We describe how \name{} obtains table context through contextualized column embeddings using the Transformer-architecture. 
Figure~\ref{fig:context_colemb} depicts how each Transformer block of the \name{} aggregates contextual information from all columns values (including dummy {\tt [CLS]} symbols and themselves) in the same table.
Specifically, this example illustrates the first Transformer layer calculates the {\em attention vector} by aggregating embeddings of other tokens based on the similarity against the second column's {\tt [CLS]} token.
Thus, an attention vector for the same symbol (e.g., {\tt George}) can be different when it appears in a different context. %
This resolves the ambiguity issue of conventional word embedding techniques such as word2vec or GloVe.

%% How Transformer works
After encoding tokens into token embeddings, a Transformer layer converts a token embedding into key (K), query (Q), and value (V) embeddings. A contextualized embedding for a token is calculated by the weighted average of value embeddings of all token embeddings, where the weights are calculated by the similarity between the query embedding and key embeddings. By having key embeddings and query embeddings separately, the model is able to calculate contextualized embeddings in an asymmetric manner. That is, the importance of {\tt Happy Feet} for {\tt George Miller}, which should be a key signal to disambiguate the person name, may not be necessarily equal to that of {\tt George Miller} for {\tt Happy Feet}. 
Furthermore, a Transformer-based model usually has multiple {\em attention heads} (e.g., 12 attention heads for the BERT base model.) Different attention heads have different parameters for K, Q, V calculation so that they can capture different characteristics of input data holistically. Finally, the output of a Transformer block is converted into the same dimension size as that of the input (e.g, 768 for BERT) so that the output of the previous Transformer block can be directly used as the input to the next Transformer block. 
The same procedure is carried out as many as the number of Transformer blocks (i.e., 12 blocks for the BERT Base model.)

%% Column embeddings
\noindent
{\bf Column representations.}
Since \name{} inserts dummy {\tt [CLS]} symbols for each column, we can consider the output embeddings of the pre-trained LM for those symbols as {\em contextualized column representations}. Note that \name{} is a {\em table-wise} model, which takes the entire table as input and thus contextualized column representations take into account table context in a holistic manner.
For column type prediction, \name{} attaches an additional dense layer followed by output layer with the size of $|\mathcal{C}_{\rm type}|$.

\noindent
{\bf Column-pair representations.} For column relation prediction, \name{} concatenates a corresponding pair of contextualized column representations as a contextualized column-pair representation.
The additional dense layer should capture combinatorial information between two column-level representations. Same as the column representations, column-pair representations are also table-wise representations.
In the experiment, we also tested a variant of \name{} that only takes a single column (a single column pair) as input for the column type (column relation) prediction task.

\rev{
More formally, given a table $T$,
the language model $\mathsf{LM}$ takes the serialized sequence
$\text{serialize}(T) = \{t_1, \dots, t_p \}$ of $p$ tokens as input
and outputs a sequence $\mathsf{LM}(T)$ where each element
$\mathsf{LM}(T)_i$ is a $d$-dimensional context-aware
embedding of the token $t_i$. 
Let $\{i_1, \dots, i_n\}$ be the indices of the inserted 
special $\mathsf{[CLS]}$ tokens. 
Let $g_{\rm type}$ be the column type prediction dense layer of dimension $d \times |\mathcal{C}_{\rm type}|$.
The column type model computes:
{
\setlength\abovedisplayskip{2pt}
\setlength\belowdisplayskip{2pt}
\begin{equation}
 \mathsf{softmax}( g_{\rm type} ( \mathsf{LM}(T)_{i_j} ) )   
\end{equation}
}
\noindent
as the predicted column type 
of the $j$-th column.
Similarly, for column relation prediction with dense layer
$g_{\rm rel}$  of dimension $2d \times |\mathcal{C}_{\rm rel}|$,
the column relation model computes:
{
\setlength\abovedisplayskip{2pt}
\setlength\belowdisplayskip{2pt}
\begin{equation}
 \mathsf{softmax}( g_{\rm rel} ( \mathsf{LM}(T)_{i_j} \oplus \mathsf{LM}(T)_{i_k} ) )   
\end{equation}
}
\noindent
as the predicted relation between the $j$-th and the $k$-th column
of table $T$. The $\oplus$ symbol denotes concatenation of two vectors.
\name{} then feeds the predictions and the groundtruth labels
into a cross entropy loss function to update the model parameters
(Line 9-10, Algorithm \ref{alg:training}).

}

\subsection{Learning from Multiple Tasks}
%% Training and Prediction
In the training phase, \name{} fine-tunes a pre-trained LM using two different training data and two different objectives. As shown in Algorithm~\ref{alg:training}, \name{} switches the task every epoch and updates the parameters for different objectives using different optimization schedulers.
This design choice enables \name{} to naturally handle imbalanced training data for different tasks. \rev{Furthermore, with a single objective function and a single optimizer, we need to carefully choose hyper-parameter(s) that balance different objective terms to create a single objective function (e.g., $\ell = \lambda \ell_1 + (1-\lambda) \ell_2$ like TCN~\cite{Wang:2021:TCN}.) With our strategy, we can avoid adjusting the hyper-parameter.}
Also, in Section~\ref{sec:analysis}, we will show that \name{} can be robustly trained with imbalanced training data.

Note that \name{} is not limited to training with just two tasks. By adding more output layers and corresponding loss functions, \name{} can be used for more than two tasks.
\rev{We also note that finding relevant tasks is challenging
as adding new tasks might not
necessarily improve the model's performance. This can be due to the tasks not sharing enough common knowledge to improve each other,
or noisy labels in training sets which can propagate among tasks.
}
Finding more relevant tasks and testing \name{} on them are part of our future work. 

\section{Evaluation} \label{sec:experiment}

\subsection{Dataset}\label{sec:datasets}
%% WikiTable dataset
We used two benchmark datasets for evaluation. The WikiTable dataset~\cite{Deng:2020:TURL} is a collection of tables collected from Wikipedia, which consists of 580,171 tables in total. 
\rev{The dataset provides both annotated column types and relations
for training and evaluation.
For column type prediction, the dataset provides 
628,254 columns from 397,098 tables annotated by 255 
column types.
For column relations, the dataset provides 62,954 column pairs 
annotated with 121 relation types
from 52,943 tables for training.
According to \cite{Deng:2020:TURL},
the type and relation labels are from FreeBase~\cite{bollacker2008freebase} and are
obtained by aggregating entity links attached to the original tables.}
For both tasks, we used the same train/valid/test splits as TURL~\cite{Deng:2020:TURL}.
Each column/column-pair allows to have more than one annotation, and thus, the task is a multi-label classification task.

%% VizNet dataset
The VizNet dataset~\cite{Zhang:2020:Sato} is a collection of WebTables, which is a subset of the original VizNet corpus~\cite{Hu:2019:CHI:VizNet}. The dataset is for the column type prediction task. The dataset has 78,733 tables, and 119,360 columns are annotated with 78 column types. 
\rev{
The dataset constructed the columns types
by mapping column headers to DBpedia 
types~\cite{Nguyen:2020:MTab4DBPedia} by a set of mapping rules.}
We used the same splits for the cross-validation to make the evaluation results directly comparable to \cite{Zhang:2020:Sato}.
Each column has only one label, and thus, the task is a multi-class classification task. 

\begin{table}[]
    \centering
    %\small
    \footnotesize
    \caption{ Dataset description.}\label{tab:datasets}    \vspace{-1em}
    \begin{tabular}{c|cccc}\hline
     Name      & \# tables & \# col & \# col types & \# col rels\\\hline\hline
     WikiTable & 580,171 & 3,230,757  & 255 & 121 \\
     VizNet & 78,733 & 119,360 & 78 & -- \\\hline
    \end{tabular}
    \vspace{-2em}
\end{table}

\subsection{Baselines}
\noindent
{\bf TURL~\cite{Deng:2020:TURL}} is a recently developed pre-trained Transformer-based LM for tables. TURL further pre-trains a pre-trained LM using table data, so the model becomes more suitable for tabular data. 
Since TURL relies on entity-linking and meta information such as table headers and table captions, which are not available in our scenario, we used a variant of TURL \rev{pre-trained on} table values for a fair comparison. \rev{Note that to perform column type/relation annotation, we fine-tuned the pre-trained TURL model on the same training sets
as for \name{} and other baselines.}

\noindent
{\bf Sherlock~\cite{Hulsebos:2019:Sherlock}} is a single-column prediction model that uses multiple feature sets, including character embeddings, word embeddings, paragraph embeddings, and column statistics (e.g., mean, std of numerical values.)
A multi-layer ``sub'' neural network is applied to each column-wise feature set to calculate compact dense vectors except for the column statistics feature set, which are already continuous values. The output of the subnetworks and the column statistics features are fed into the ``primary'' neural network that consists of two fully connected layers. 

\noindent
{\bf Sato~\cite{Zhang:2020:Sato}} is a multi-column prediction model, which extends Sherlock by adding LDA features to capture table context and a CRF layer to incorporate column type dependency into prediction.
Sato is the state-of-the-art column type prediction on the VizNet dataset.

\subsection{Experimental Settings}
We used Adam optimizer with an $\epsilon$ of $10^{-8}$. The initial learning rate was set to be $5\times 10^{-5}$ with a linear decay scheduler with no warm-up.
We trained \name{} for 30 epochs \rev{and chose the checkpoint with the highest F1 score on the validation set.}

%% BCELoss and Cross-entropy loss
Since the WikiTable dataset can have multiple labels on each column/column pair, we used Binary Cross Entropy loss to formulate as a multi-label prediction task.
For the VizNet dataset, which only has a single annotation on each column, we used Cross Entropy loss to formulate as a multi-class prediction task.
Models and experiments were implemented with PyTorch~\cite{PyTorch:NeurIPS2019} and the Transformers library~\cite{Wolf-etal:2020:Transformers}. All experiments were conducted on an AWS {\tt p3.8xlarge} instance (V100 (16GB)).

Following the previous studies~\cite{Deng:2020:TURL,Zhang:2020:Sato}, we use micro F1 for the WikiTable dataset, and micro F1 and macro F1 for the VizNet dataset, as evaluation metrics. 
\rev{The micro F1 score is the weighted average of F1 values based on the sample size of each class, while the macro F1 score is the simple average of F1 values for all classes.}
\ifnum \arxiv=1
\revcr{
Additional results and analysis can be found in Appendix.
}
\else
\revcr{
Additional results and analysis can be found in the appendix of the full version \cite{suhara2021annotating}.
}
\fi

\begin{table}[t]
\centering
%\small
\footnotesize
\caption{Performance on the WikiTable dataset.
}\label{tab:main_turl}
\vspace{-1em}
\begin{tabular}{c|ccc|ccc}\hline
Method & \multicolumn{3}{c|}{Col type} & \multicolumn{3}{c}{Col rel} \\
       & P & R & F1 & P & R & F1 \\\hline\hline
Sherlock
& \rev{88.40} & \rev{70.55} & 78.47 & -- & -- & --\\
TURL     & \rev{90.54} & \rev{87.23} & 88.86 & \rev{91.18} & \rev{90.69} & 90.94 \\
\name   & \rev{{\bf 92.69}} & \rev{{\bf 92.21}} & {\bf 92.45} & \rev{{\bf 91.97}} & \rev{{\bf 91.47}} & {\bf 91.72} \\\hline
\rev{TURL+metadata} & \rev{92.75} & \rev{92.63} & \rev{92.69} & \rev{92.90} & \rev{93.80} & \rev{93.35} \\
\rev{\name+metadata} & 93.25 & 92.34 & 92.79 & 91.20 & 94.50 & 92.82 \\\hline
\end{tabular}
%    \vspace{-1em}
\end{table}

\subsection{Main Results}\label{sec:mainresult}
%% Main results
\noindent
{\bf WikiTable} Table~\ref{tab:main_turl} shows the micro F1 performance for the column type prediction and column relation prediction tasks on the WikiTable dataset.
\name{} significantly outperforms the state-of-the-art method TURL on both of the tasks \hl{with improvements of 4.0\% and 0.9\%, respectively.}

\hl{
A significant difference in the model architecture between \name{} and TURL is whether the model uses full self-attention. In TURL, the model uses the self-attention mechanism with the ``cross-column'' edges removed, which they referred to as visibility matrix~\cite{Deng:2020:TURL}.
Let us use the example in Figure~\ref{fig:context_colemb}, which depicts how the contextualized embedding for the second column is calculated.  
TURL's visibility matrix removes the connections to ${\tt [CLS]}^2$ from 
the cells \texttt{``Happy Feet'', ``Cars'', ``USA'', and ``UK''},
whereas our \name{} uses the full set of connections.}

Since TURL is designed for tables with meta information (e.g., table captions or column headers), we consider the major benefit of this design (i.e., the visibility matrix) to effectively incorporate descriptions in the meta information into table values. 
From the results, \name{} with the full self-attention performs better than TURL, which indicates that some direct intersections between tokens in different columns and different rows are useful for the column annotation problem.

%% Meta information
\rev{We also tested \name{} with metadata, which appends the column name to column values for each column before serialization.  As shown in Table~\ref{tab:main_turl}, by using column names, \name{} slightly improves the performance and performs competitively against TURL with metadata. This indicates that TURL relies on metadata and \name{} performs better and more robustly than TURL when metadata is not available.}

\iffalse
\begin{table}[t]
\centering
\caption{Performance on the WikiTable dataset.\jinfeng{Suggest to print out F1 scores on individual types to find out what kind of types DODUO outperforms the baselines significantly.}}\label{tab:main_turl}
\begin{tabular}{c|cc|cc}\hline
 & \multicolumn{2}{c|}{Type prediction} & \multicolumn{2}{c}{Relation prediction} \\
 Method & Macro F1 & Micro F1 & Macro F1 & Micro F1 \\\hline\hline
Sherlock &  -- & 78.47 & -- & -- \\
TURL     &  -- & 88.86 & -- & 90.94 \\\hline
\name    &  -- & {\bf 92.50} & {\bf 85.85} & {\bf 91.90} \\\hline
\end{tabular}
\end{table}
\fi

\begin{table}[t]
\centering
%\small
\footnotesize
\caption{Performance on the VizNet dataset.}\label{tab:sato_main}
\vspace{-1em}
\begin{tabular}{c|cc|cc}\hline
 & \multicolumn{2}{c|}{Full} & \multicolumn{2}{c}{Multi-column only} \\
 Method & Macro F1 & Micro F1 & Macro F1 & Micro F1 \\\hline\hline
Sherlock &  69.2 & 86.7 & 64.2 & 87.9\\
Sato & 75.6 & 88.4 & 73.5 & 92.5\\
\name  & {\bf 84.6} & {\bf 94.3} & {\bf 83.8} & {\bf 96.4} \\\hline
\end{tabular}
    \vspace{-1em}
\end{table}

\noindent
{\bf VizNet} Table~\ref{tab:sato_main} shows the results on the VizNet dataset. Note that \name{} is trained only using the column prediction task for the VizNet dataset, as column relation labels are not available for the dataset.
The results show that \name{} outperforms Sherlock and Sato, the SoTA method for the dataset, by a large margin and establishes new state-of-the-art performance with micro F1 (macro F1) improvements of 11.9\% (6.7\%.)

%% Difference against Sato
As described in Section~\ref{sec:related},
Sato is a multi-column model that incorporates table context by using LDA features.
Different from the LDA features that provide multi-dimensional vector representations for the entire table, the Transformer-based architecture enables \name{} to capture more fine-grained inter-token relationships through the self-attention mechanism. Furthermore, \name's table-wise design naturally helps incorporate inter-column information into the model.

\rev{
\smallskip
\noindent
\textbf{Performance on numeric columns. }
As mentioned in Section \ref{sec:background},
\name{} casts all cell values as strings thus
it may be weak in handling numeric columns.
Table~\ref{tab:numeric} shows \name{} performance
on the top-15 most numeric column types 
from the VizNet dataset.
\name{} did have low performance on some numeric types
such as ranking (33.21\%) and capacity (62.55\%).
However, on these 15 types, \name{} achieves an average
F1 score of 86.9\% which is 
comparable to the overall macro F1 (84.6\%) and
even slightly better. 
This can be due to the Transformer model
being able to recognize digit patterns in the
numeric values to predict the correct types. 
This results aligns with the findings from the 
NLP literature~\cite{wallace2019nlp,geva2020injecting}
that Transformer models can partially handle
numeric data.

\begin{table}[!ht]
\caption{\rev{\name{}'s column type prediction performance on the 15 most
numeric types of the VizNet dataset. We use \%num to measure how many cell values of a type can be cast as a numeric type (e.g., \texttt{int}, \texttt{float}, \texttt{date}).}}\vspace{-1em}
\label{tab:numeric}
\resizebox{0.48\textwidth}{!}{  
\begin{tabular}{ccccccccc}\toprule
type  & \%num  & F1    & type      & \%num & F1    & type     & \%num & F1    \\ \midrule
plays & 100.00 & 88.55 & fileSize  & 87.84 & 88.23 & grades   & 67.18 & 97.68 \\
rank  & 93.01  & 94.52 & elevation & 87.39 & 92.14 & weight   & 60.41 & 97.59 \\
depth & 92.86  & 88.45 & ranking   & 86.88 & 33.21 & isbn     & 43.77 & 96.51 \\
sales & 92.05  & 75.13 & age       & 81.04 & 98.53 & capacity & 42.06 & 62.55 \\
year  & 91.47  & 98.94 & birthDate & 67.85 & 95.64 & code     & 35.93 & 95.43 \\ \bottomrule
\end{tabular}}
\end{table}
}

\section{Analysis}\label{sec:analysis}

\begin{table}[t]
\centering
%\small
\footnotesize
\caption{Ablation study on the WikiTable dataset.}\label{tab:ablation_turl}
\vspace{-1em}
\begin{tabular}{l|c|c}\hline
Method & Type prediction & Relation prediction \\\hline\hline
\name    &  {\bf 92.50} & {\bf 91.90} \\
\quad \rev{w/ shuffled rows} & \rev{91.94} & \rev{91.61} \\
\quad \rev{w/ shuffled cols} & \rev{92.68} & \rev{91.98} \\\hline
\namest & 91.37 (1.23\% $\downarrow$) & 91.24 (0.7\% $\downarrow$) \\
\namesc & 82.45 (21.9\% $\downarrow$) & 83.08 (9.6\% $\downarrow$) \\\hline
\end{tabular}
\vspace{-.5em}
\end{table}

\begin{table}[t]
\centering
%\small
\footnotesize
\caption{Ablation study on the VizNet dataset (Full.)}\label{tab:ablation_sato}
\vspace{-1em}
\begin{tabular}{c|c|c}\hline
 & Macro F1 & Micro F1  \\\hline\hline
\name    &  {\bf 84.6} & {\bf 94.3} \\
\namesc & 77.4 (8.5\% $\downarrow$) & 90.2 (4.3\% $\downarrow$) \\\hline
\end{tabular}
\vspace{-1em}
\end{table}

\subsection{Ablation Analysis}\label{subsec:ablation}
To verify the effectiveness of multi-task learning and multi-column architecture, we tested variants of \name{}.
\namest{} is a \name{} model without multi-task learning. Thus, we trained \name{} models only using training data for the target task (i.e., column type prediction or column relation prediction.)
\namesc{} is a single column model that only uses column values of the target column (or target column pair for column relation prediction.) \namesc{} is also trained without multi-task learning.

Table~\ref{tab:ablation_turl} shows the results of the ablation study. For both of the tasks, \namest{} degraded the performance compared to the multi-task learning of \name. 
As \namesc{} shows significantly lower performance than the others, the results also confirm that the multi-column architecture of \name{} successfully captures table context to improve the performance on the column type prediction and column relation prediction tasks.

The same analysis with \namesc{} on the VizNet dataset is shown in Table~\ref{tab:ablation_sato}.
As expected, the multi-column model (\name) performs significantly better than the single-column model (\namesc.) The results further confirm the strengths of the multi-column model. We would like to emphasize that \namesc{} outperforms Sato, which incorporates table context as LDA features.

\rev{The pre-trained LM (e.g., BERT) is sensitive to the input sequence order, which may not reflect the property of tables that rows/columns are order-invariant. To verify if \name{} has this limitation, we trained and evaluated \name{} on two versions of the WikiTable dataset, where the input table's rows (columns) were randomly shuffled. As shown in Table~\ref{tab:ablation_turl}, somewhat surprisingly, \name{} shows only subtle degradation for shuffled rows and no substantial difference for shuffled columns. We conjecture that \name{} successfully tailors the original position embeddings to be aligned with the table structure during the fine-tuning step.}

\subsection{Data Efficiency}
\noindent
{\bf Learning Efficiency} 
\hll{Pre-trained LMs are known for its capability of training high-quality models using a relatively small number of labeled examples.}
Furthermore, multi-task learning (i.e., training with \hll{multiple tasks simultaneously})
should further stabilize the performance with fewer training data for each task.
To verify the effectiveness of \name{} \rev{and \namest{}} with respect to \hll{label efficiency,}
we compared the \name{} models trained with different training data sizes (10\%, 25\%, 50\%, and 100\%) and evaluated the performance on the column type prediction and column relation prediction tasks. 

\begin{figure}[t]
    \centering
     \begin{subfigure}[b]{0.23\textwidth}
         \centering
         \includegraphics[width=\textwidth]{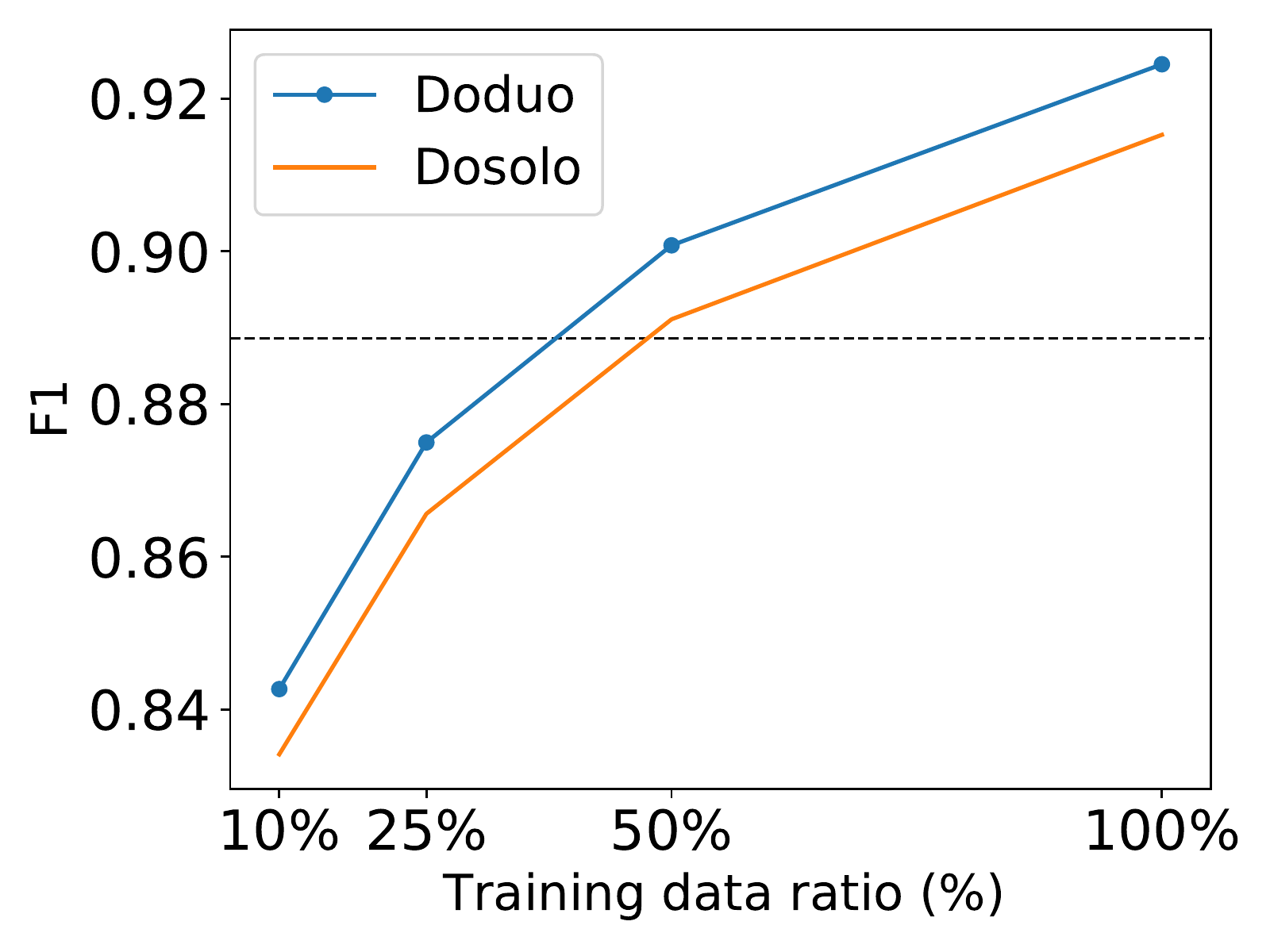}
         \caption{Column type prediction}
     \end{subfigure}
     \hfill
     \begin{subfigure}[b]{0.23\textwidth}
         \centering
         \includegraphics[width=\textwidth]{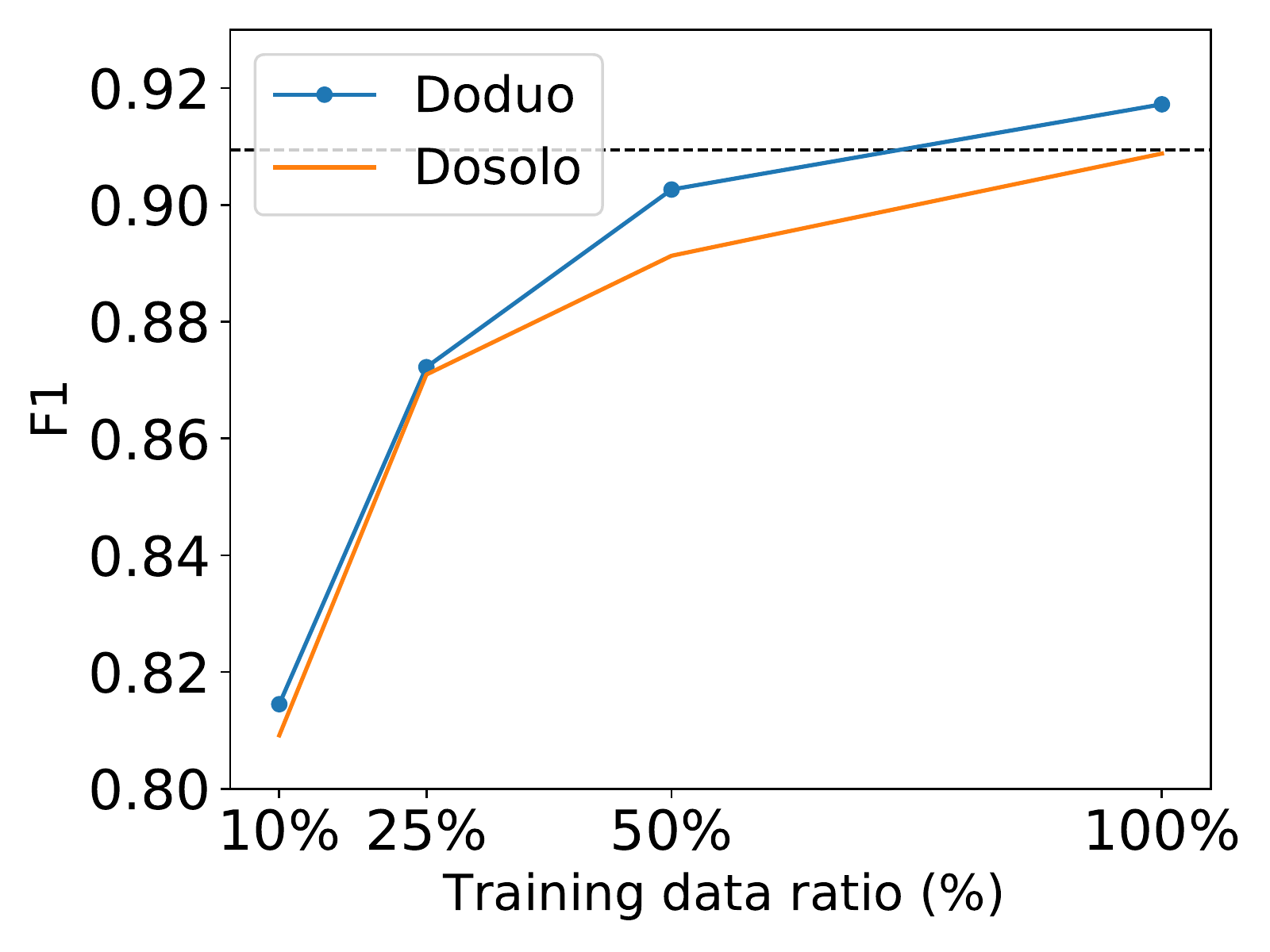}
         \caption{Column relation prediction}
     \end{subfigure}
     \vspace{-0.5em}
     \caption{\rev{Performance improvements} over increasing the training data size 
     on the WikiTable dataset. The dashed lines in the plots denote the state-of-the-art methods (TURL.) 
     }\label{fig:lc}
     \vspace{-1em}
\end{figure}

As shown in Figure~\ref{fig:lc}, \name{} \rev{consistently outperforms \namest{} and}
achieves higher than 0.9 F1 scores on both tasks even when trained with half of the training data.
\hll{In particular, with only 50\% or fewer labeled examples, \name{} outperforms the SoTA method TURL on column-type prediction and achieves comparable performance on column relation prediction.}

\begin{table}[t]
\centering
%\small
\footnotesize
\caption{Comparisons of \name{} with different input token size on the WikiTable dataset.}\label{tab:turl_maxtoken}
\vspace{-1em}
\begin{tabular}{c|c|c||c}\hline
MaxToken/col & Col type (F1) & Col rel (F1) & Max. \# of cols\\\hline\hline
8 & 89.8 & 88.9 & 56\\
16 & 91.4 & 90.7 & 30\\
32 & {\bf 92.4} & {\bf 91.7} & 15\\ \hline
\end{tabular}
\vspace{-1em}
\end{table}

\smallskip
\noindent
{\bf Input Data Efficiency} \hll{A major challenge of applying pre-trained LMs to data management tasks is their limits of the maximum input sequence length. For example, LMs like BERT can only take at most 512 tokens so ingesting a full wide table 
may not be feasible for the LMs.
\name{} (or the multi-column model in general) has the advantage that 
it is \emph{input data efficient}, meaning that it can make table-wise predictions accurately by only taking a small number of samples of each column. This makes \name{} more attractive in practice as it can handle large tables with many columns.}

Thus, we evaluated different variants of \name{} with shorter input token length to discuss the 
\hll{input} data efficiency of \name. We would like to emphasize that any of the recent studies applying pre-trained Transformer-based LMs to data management tasks (e.g., \cite{Li:2020:Ditto,Deng:2020:TURL,Trabelsi:2020:SeLaB,Wang:2021:TCN}) did not conduct this kind of analysis. Thus, it is still not clear how many tokens should we feed to the model to obtain reasonable task performance.

%% WikiTable
Table~\ref{tab:turl_maxtoken} shows the results of \name{} with different max token sizes on the WikiTable dataset. 
We simply truncated column values if the number of tokens exceeded the threshold.
As shown in the table, the more tokens used, the better performance \name can achieve, as expected. However, somewhat surprisingly, \name{} already outperforms TURL using just 8 tokens per column for column type prediction (TURL has micro F1 of 88.86 on WikiTable).
For the column relation prediction task, \name{} needs to use more tokens to outperform TURL (i.e., 32 tokens to beat TURL's score of 90.94.) This is understandable, as column relation prediction is more contextual than column type prediction, and thus it requires more signals to further narrow down to the correct prediction.
For the VizNet dataset, we confirm that \name{} with 8 max tokens per column with \name{} (92.5 F1) outperforms the state-of-the-art method (i.e., Sato) (88.4 F1) on the task. \hlll{Table~\ref{tab:turl_maxtoken} reports how many columns each variant can support under the maximum token configuration.
The average numbers of columns in Web Tables, Enterprise Data, and Open Data are reported to be 4, 12, and 16, respectively~\cite{Miller:2018:VLDB:OpenDataIntegration,Deng:2018:CIDR}.
Thus, we confirm that \name{} has a nice input data efficiency property, so it can be also used for ``wide'' tables.}

\rev{
We note that 16 columns may not be sufficient for tables outside of WebTables. 
For such cases, a reasonable option is to first split the wide table into clusters of relevant columns (maybe by some user-defined rules), then apply \name{} on each cluster. 
In this case, \name{} still has the advantage 
of leveraging partial context of the input table
to improve prediction quality.}

\section{Case Study: Clustering Columns}
\hll{
We apply \name{} to a real data science application scenario of clustering relevant columns.
On a daily basis, data scientists collect information from multiple sources and incorporate data from various tables. Therefore, as a first step of data exploration and data integration~\cite{Ota:2020:D4:VLDB}, it is essential to know which columns are semantically similar. However, this is not always straightforward as different column names can be assigned to semantically similar columns. 
In this section, we demonstrate the usefulness of \name's contextualized column embeddings with a case study of clustering semantically similar columns on an enterprise database.

% setting of the scenario
For this case study, we use an in-production enterprise database from the HR domain. Different teams create and update tables about job seekers and companies that are hiring, etc. Despite some company-wide naming conventions, we observe that semantically similar columns are sometimes given different column names across tables. 
Some tables have additional meta information describing the meaning of each column, which helps to understand the similarity of columns. However, the meta information is missing for many columns and is not always reliable as they are worded differently  by different teams.
Next, we simulate a workflow of a data scientist performing analysis related to job search and review of companies.

\smallskip
\noindent
{\bf Scenario:}
Our data scientist \hlll{Sofia} starts by filtering tables with keywords ``jobsearch'' and ``review'' and gets {\bf 10} tables with {\bf 50} columns in total ({\bf 29} columns of type ``string'' and {\bf 21} columns of type ``integer''.)
To group similar columns together, she can simply create contextualized column embedding using the \name{} toolbox for all the columns and then apply her favorite clustering algorithm to form the final groups.

% Evaluation
%ground-truth
\hlll{To evaluate Sofia's column groups}, we generate an initial cluster using both the column names and descriptions and manually refine the results to form the ground-truth\footnote{We used column name and description for ground-truth creation purpose only as it is not available for all tables in practice. The initial clustering uses a combination of TF-IDF vectors and fastText embeddings.} \revcr{as shown below:}
{
\begin{table}[h]
\footnotesize
\begin{tabular}{|p{8cm}|}\hline
date, IP address, job title, timestamp (unixtime), timestamp (hhmm), counts, status, file path, browser, location, search term, rating, company ID, review ID, user ID \\\hline
\end{tabular}
\end{table}
}

Sofia use the \name{} model trained on the WikiTable dataset to obtain contextualized column embeddings for each column (\name + column value emb.) 

We compared the method with three baselines and \rev{two traditional schema matching approaches}. We directly used \name's column type predictions as the clustering criteria where columns with the same predicted types got assigned into the same cluster (\name+predicted type).
We tested fastText~\cite{bojanowski2016enriching} to verify how {\em non}-contextualized column embeddings perform for the task. We use column value embeddings (fastText+column value emb) and column name embeddings (fastText+column name emb) with fastText. We choose fastText as a baseline as it offers a widely used off-the-shelf toolbox, which is a ``go-to'' option for data scientists. 
To achieve a fair evaluation of the embedding quality, we use the same k-means clustering algorithm for all models.
\rev{In addition, we compared with more traditional schema matching approaches tested in the experiment suite Valentine \cite{koutras2021valentine}. We picked the two most effective approaches from Valentine's empirical study: COMA~\cite{do2002coma} and DistributionBased~\cite{zhang2011automatic}. To generate the clustering label for comparison, we went over all possible pairs of tables and connected matched columns to assign the same cluster labels.} 
\revcr{Since those schema matching methods take two tables as input and return pairs of matched columns, we regard returned pairs as connected nodes in a graph and merge them into connected components to obtain column clusters.}
We use Homogeneity (Precision), Completeness (Recall), V-Measure (F1) to evaluate the quality of clusters using the ground-truth cluster assignment.

%% Results
As shown in Table~\ref{tab:case_study}, \name's column embeddings show the best clustering performance with respect to Precision and F1. The results confirm that contextualized column embeddings are more useful than predicted column types and can be more accurate representations than column name/value embeddings created by fastText.
Compared to \name, fastText tends to generate similar embeddings even for semantically different columns, which leads to creating unnecessarily large clusters that contain many irrelevant columns. This significantly increases (decreases) fastText methods' Recall (Precision) values in Table~\ref{tab:case_study}.
\rev{COMA shows reasonably good performance and DistributionBased falls short on precision, given both column name and content. \name{} outperforms both matching-based approaches using the contextualized embedding.}
Note that both the column names and column descriptions are not given to the \name{} model as input, and the model was trained on the WikiTable dataset (i.e., different domain.)
Thus, this case study also indicates the transferability of a \name{} model trained on one domain (i.e., Wikipedia tables) to another domain (i.e., enterprise database.) so that data scientists can apply the \name{} model in the toolbox for their own needs.}

\begin{table}[t]
\centering
%\small
\footnotesize
\caption{Case study results.}\label{tab:case_study} 
\vspace{-1em}
\begin{tabular}{c|ccc}\hline
Method & Prec. & Recall & F1 \\\hline\hline
\name+column value emb & {\bf 68.19} & 70.40 & {\bf 69.28} \\\hline
\name+predicted type & 44.87 & 61.32 & 51.82 \\
fastText+column value emb & 35.90 & {\bf 76.61} & 48.89 \\
fastText+column name emb & 56.62 & 74.68 & 64.40 \\\hline 
\rev{COMA (with column name)} & \rev{58.47} & \rev{66.06} & \rev{62.03} \\
\rev{DistributionBased (with column name)} & \rev{23.87} & \rev{69.51} & \rev{35.53} \\\hline
\end{tabular}
\vspace{-1em}
\end{table}

\section{Conclusion} \label{sec:conclude}
In this paper, we present \name{}, a unified column annotation framework based on pre-trained Transformer language models and Multi-task learning.
Experiments on two benchmark datasets show that \name{} achieves new state-of-the-performance. With a series of analyses, we confirm that the improvements are benefited from the multi-task learning framework.
Through the analysis, we also confirm that \name{} is data-efficient, as it can achieve competitive performance as the previous state-of-the-art methods only using 8 tokens per column or about 50\% of training data.
\hll{We conduct a case study and verify the effectiveness of \name{} on a real-world data science problem. We believe our toolbox will further help researcher/data scientists easily apply the state-of-the-art column annotation model to a wide variety of data science and data management problems.}

\ifnum \arxiv=1

\begin{acks}
We thank Jin Wang and Estevam Hruschka for their valuable feedback  on the draft. 
\end{acks}
\fi

%\clearpage

%\newpage
\balance
\bibliographystyle{ACM-Reference-Format}
\bibliography{refs}

\ifnum \arxiv=1

\appendix
\section{Additional Analysis}

\subsection{Detailed Results}
%% Class F1 scores
Figure~\ref{fig:class_f1_sato} shows the F1 score of \name{} and Sato for each class on the VizNet (Full) and VizNet (Multi-column only) datasets. 
The improvements against Sato on the two variants of the VizNet datasets indicate that \name{} robustly and consistently performs better than Sato, especially for single-column tables, where Sato cannot benefit from the table features and CRF.
We found that Sato shows zero or very poor F1 values for {\tt religion}, {\tt education}, {\tt organisation}.
\hl{The labeled columns in the training data in the 1st fold of the VizNet (Full) are only 24, 22, and 14, respectively. Sato should suffer from the lack of training examples for those column types, and probably the skewed column type distribution as well. We show that \name{} robustly performs well on such column types.}

Table~\ref{tab:tur_class_f1} shows the performance of \name{} and \namest{} on the WikiTable dataset for 6 column types/column relations.
We confirm that \name{} tends to perform better for the column types/relations that are less clearly distinguishable (e.g., artist vs. writer, place-of-birth vs. place-lived.)

\begin{figure*}[t]
         \centering
         \begin{subfigure}[b]{0.95\textwidth}
             \includegraphics[width=\textwidth]{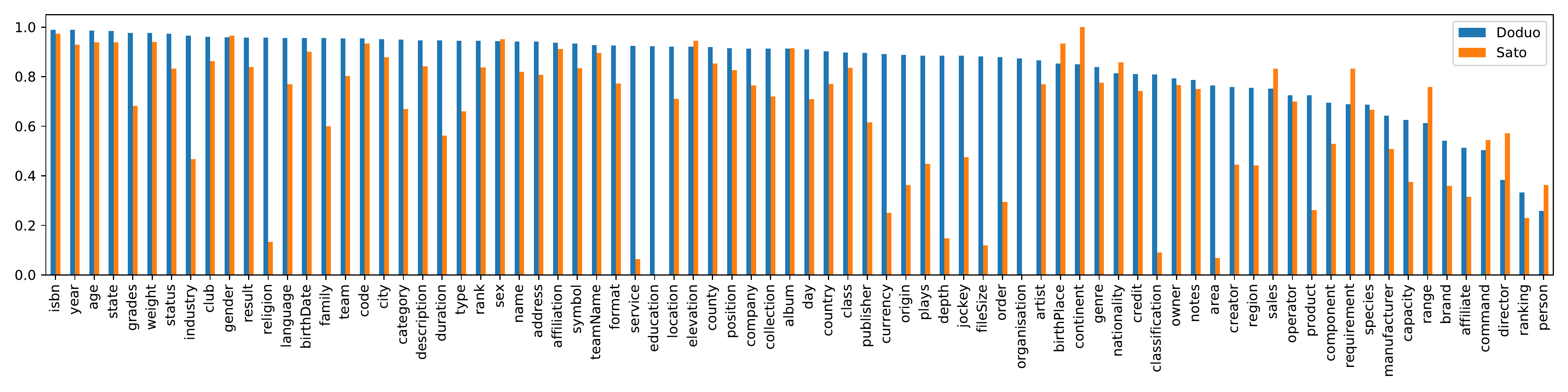}
         \end{subfigure}
         \begin{subfigure}[b]{0.95\textwidth}
             \includegraphics[width=\textwidth]{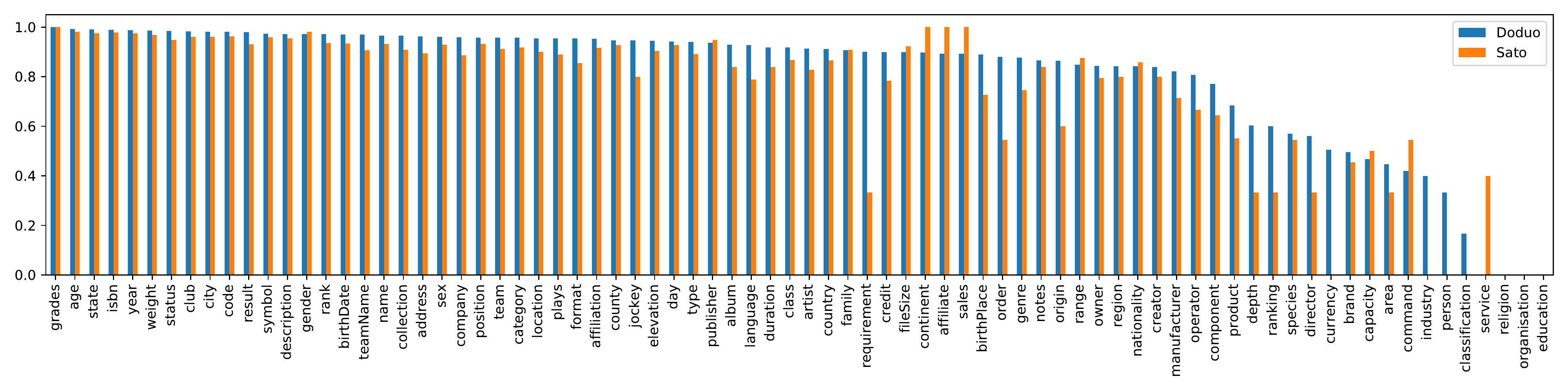}
         \end{subfigure}
         \caption{Class F1 values by \name{} and Sato on the VizNet dataset (Above: Full set; Below: Multi-column only.)}\label{fig:class_f1_sato}
\end{figure*}

\begin{table*}[t]
\caption{Further analysis on column type prediction (left) and on column relation prediction (right.)}\label{tab:tur_class_f1}
\begin{subfigure}[b]{0.49\textwidth}
    \centering
    %\footnotesize
    \small
    \begin{tabular}{p{5.1cm}|c|c}\hline
    Column type & \name{} (F1) & \namest{} (F1) \\\hline\hline
    {\tt music.artist}                          &   {\bf 84.03}  &   81.87 \\
    {\tt music.genre}                           &   {\bf 93.33}  &   87.50 \\
    {\tt music.writer}                          &   {\bf 75.00}  &   40.00 \\ 
    {\tt american\_football.football\_coach}      &  {\bf 70.59}  &   66.67 \\
    {\tt american\_football.football\_conference} &  {\bf 44.44}  &   36.36 \\
    {\tt american\_football.football\_team}       &  {\bf 86.67}  &   86.36 \\\hline
    \end{tabular}
\end{subfigure}
\begin{subfigure}[b]{0.49\textwidth}
    \centering
    %footnotesize
    \small
    \begin{tabular}{p{4.1cm}|c|c}\hline
    Column relation & \name{} (F1) & \namest{} (F1) \\\hline\hline
    {\tt film.film.production\_companies}                          & {\bf 80.95}  & 74.29 \\
    {\tt film.film.produced\_by}                                   & {\bf 43.90}  & 38.89	\\
    {\tt film.film.story\_by}                                      & {\bf 100.00} & 90.91	\\
    {\tt people.person.place\_of\_birth}                           & {\bf 92.00}  & 90.79	\\
    {\tt people.person.place\_lived} & {\bf 85.98}  & 77.67	 \\
    {\tt people.person.nationality}                                & {\bf 100.00} & 98.80 \\\hline
    \end{tabular}
\end{subfigure}
\end{table*}

\subsection{Input Token Length}

%% VizNet
For the VizNet dataset, we tested \name{} and \namesc{} with different maximum token numbers per column. 
Table~\ref{tab:sato_maxtoken} shows the similar trends as the results on the WikiTable dataset. \name{} with 8 max tokens per column with \name{} outperforms the state-of-the-art method (i.e., Sato) on the task.
As we observe significant differences between the multi-column model (\name) and the single-column model (\namesc), we consider it is mainly because the Transformer blocks (i.e., self-attention mechanisms) capture the inter-column table context successfully.

\begin{table}[t]
\centering
\caption{Comparisons with different input token size on the VizNet (Full) dataset.}\label{tab:sato_maxtoken}
\begin{tabular}{c|c|cc}\hline
 Method & MaxToken/col & Macro F1 & Micro F1 \\\hline\hline
\name      & 8 & 81.0 & 92.5 \\
           & 16 & {\bf 83.6} & 93.6 \\
           & 32 & 83.4 & {\bf 94.2} \\\hline
\namesc    & 8 & 72.7 & 87.2 \\
           & 16 & 76.1 & 89.1 \\
           & 32 & 77.4 & 90.2 \\\hline
\end{tabular}
\end{table}

\subsection{Learning Efficiency}

\subsection{\hl{Inter-column Dependency}}\label{subsec:attention}
%% Overview
A strength of the Transformer architecture is stacked Transformer blocks that calculate highly contextual information through the self-attention mechanism.
As described in Section~\ref{sec:model}, \name{} uses the {\tt [CLS]} dummy symbols to explicitly obtain contextualized column representations. The representations not only take into account table context but also explicitly incorporate the inter-column dependency. That is, as we showed in Figure~\ref{fig:examples}, predictions for some columns should be relevant and useful for other columns in the table.
To the best of our knowledge, none of the existing work that applies pre-trained LMs to tables has conducted this type of analysis for better understanding how the Transformer-based model captures the semantics of tables.

Thus, we conduct attention analysis to further understand how the attention mechanisms of \name{} (i.e., the pre-trained LM) captures the inter-column dependency and the semantic similarity between them.
Following the literature of attention analysis in NLP~\cite{Clark:2019:WhatDoesBERTLookAt,tenney-etal-2019-bert-rediscovers,Rogers:2020:TACL:BERTology}, we look into attention weights for the analysis.
It is known that in pre-trained Transformer-based LMs, 
the deep layer focuses on semantic similarity between tokens~\cite{Clark:2019:WhatDoesBERTLookAt,tenney-etal-2019-bert-rediscovers}.
Therefore, to investigate the high-level (semantic) similarity between columns, we looked into the attention weights of the last Transformer block. 
%

%% Details
We used the VizNet dataset (Multi-column only) for the analysis.
Specifically, we focus on attention weights between {\tt [CLS]} tokens (i.e., column representations.) 
Since Transformer-based LMs usually have multiple attention heads (e.g., 12 heads in the BERT Base model,) we aggregate attention weights of all attention heads. As a result, we obtain an $S \times S$ matrix, where $S$ denotes the input sequence length.
We disregard aggregated attention weights other than those for {\tt [CLS]} tokens.
\hl{After masking out any attention weights other than {\tt [CLS]} tokens, we averaged the matrices obtained from all tables in the dataset so that we can create a single $|\mathcal{C}_{\rm type}| \times |\mathcal{C}_{\rm type}|$ matrix that represents the dependency between column types. This gives us aggregated information about the dependency between column types.}

Each element ($i$, $j$) in the final matrix represents how much the column type $i$ relies on the other column type $j$ for its contextualized representation.
Note that the \hl{dependency} of column type $i$ (or $j$) for column type $j$ (or $i$) can be different, and thus the matrix is not symmetric. \hl{For example, {\tt age} highly relies on the type {\tt origin}, whereas the opposite direction has negative attention weight showing a low degree of dependency.}
To eliminate the influence of the co-occurrence of column types, we counted the co-occurrence of column types in the same table and \hl{normalized the matrix to make the reference point to be zero for more straightforward interpretation.
As a result, the final matrix consists of relative importance scores, and higher/lower values mean more/less influence from the column type.}

\hl{Figure~\ref{fig:attention} depicts the final matrix in heatmap visualization. Higher values (colored in red) indicate stronger dependency of column types (in $y$-axis) against other column types (in $x$-axis). For example, {\tt gender} ({\tt age}) has a higher value against {\tt country} ({\tt origin}.) We can interpret that majority of information, which composes the contextualized column representations for {\tt gender} columns, is derived from {\tt origin.} On the other hand, the {\tt gender} column seems not to be important for the {\tt origin} column.}
The results confirm that \name{} learns the inter-column dependency through the self-attention mechanism and the learned semantic similarity values between different pairs of column types have different weights, which the co-occurrence cannot simply explain.

\begin{figure}[t]
    \centering
    \includegraphics[width=0.50\textwidth]{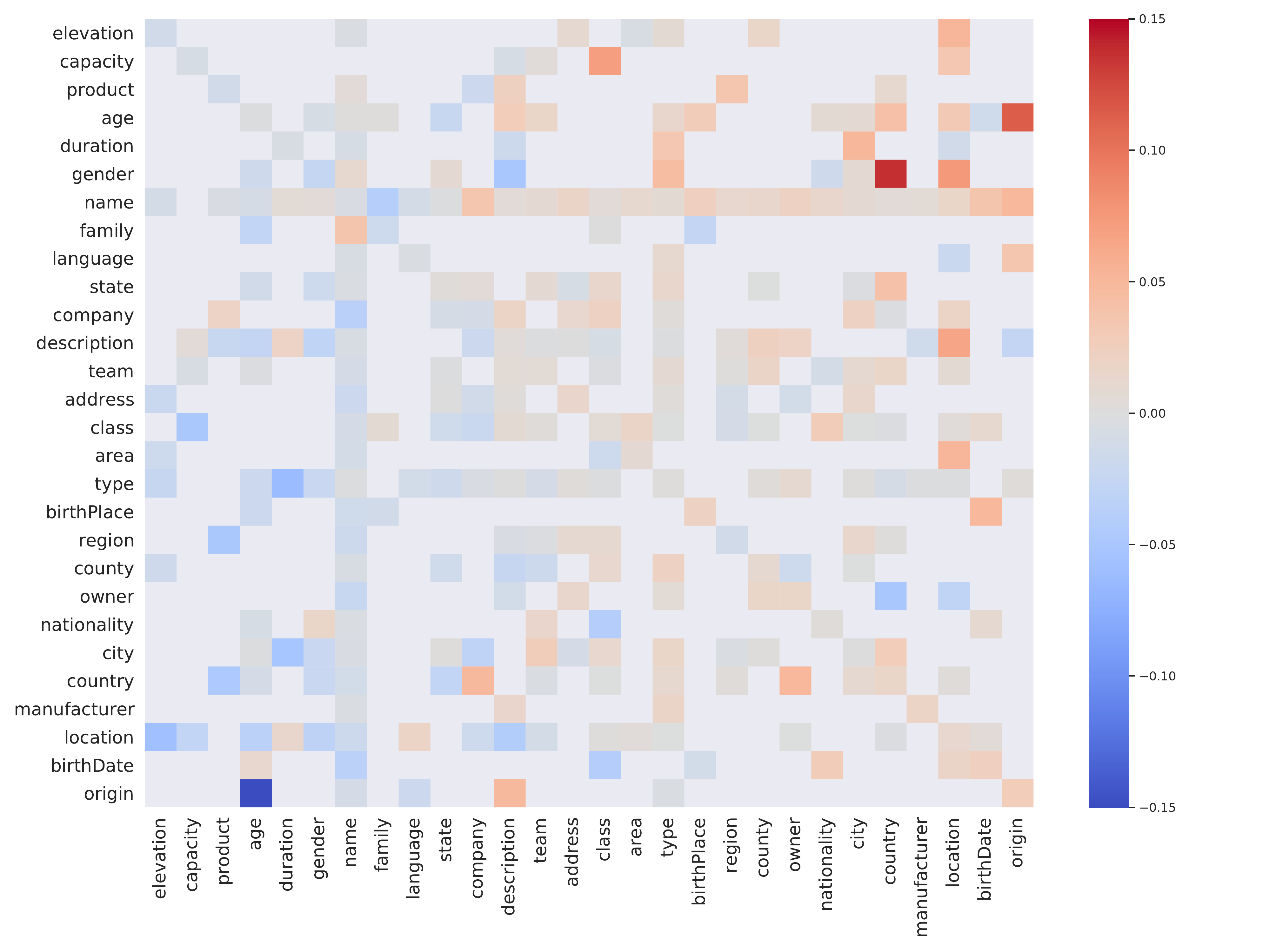}
    \caption{Inter-column dependency based on attention analysis on the VizNet dataset. A higher value (red) indicates that the column type ($y$-axis) ``relies on'' the other column type ($x$-axis) for prediction. Each row denotes the degree of ``dependence'' against each column. For example, {\tt age} in $y$-axis has a high attention weight against {\tt origin} in $x$-axis, indicating that predicting {\tt age} relies on signals from the {\tt origin} column.}
    \label{fig:attention}
    \vspace{-0.5cm}
\end{figure}

\subsection{\hl{Language Model Probing}}\label{subsec:lm_probing}
Recent applications of pre-trained LMs to data management tasks including entity matching~\cite{Brunner:2020:EDBT:BERT4EM,Li:2020:Ditto} and column annotation tasks~\cite{Deng:2020:TURL,Wang:2021:TCN} have shown great success by improving previous SoTA performance by a large margin.
However, little is known about how well pre-trained LMs {\em inherently} know about the problem in the first place. 
Pre-trained LMs are trained on large-scale textual corpora, and the pre-training process helps the pre-trained LM to memorize and generalize the knowledge stored in the pre-training corpus.

% why we did this analysis
There is a line of recent work that aims to investigate how well pre-trained LMs know about factual knowledge~\cite{Petroni:2019:LAMA,Jiang:2020:HowCanWeKnowLM,Roberts:2020:HowMuchKnowledge}. The studies have shown that pre-trained LMs store a significant amount of factual knowledge through pre-training on large-scale corpora.
Therefore, we hypothesized that \name's performance was partly boosted by the knowledge obtained from the pre-training corpus that might store knowledge relevant to the task.
To verify the hypothesis, we evaluated if the BERT model, which we used as the base model for the experiments, stored relevant knowledge for the column annotation problem.

In the analysis, following the line of work~\cite{Petroni:2019:LAMA,Jiang:2020:HowCanWeKnowLM,Roberts:2020:HowMuchKnowledge}, we use the template-based approach to test if a pre-trained LM knows the factual knowledge. 
Specifically, we use a template that has a blank field for the column type like below:
\begin{center}
    Judy Morris is \_\_\_\_\_.
\end{center}
In this example, ``director'' should be a better fit than other column types (e.g., ``actor'', ``player'', etc.)

In this way, we conclude that the model knows the fact if the model judges the template with the true column type (e.g., ``director'') {\em more likely} than other sentences that use different column types (e.g., ``actor'', ``player'', etc.).
To evaluate the likelihood of a sequence after filling the blank in the template, we use the {\em perplexity} score of a sequence using the pre-trained LM.
Perplexity is used to measure how well the LM can predict each token from the context (i.e., the other tokens in the sequence\footnote{It would be the next token from the previous tokens if the model were an autoregressive model that only considers backward dependency in each step (e.g., GPT-2.) Since BERT has bi-directional connections between any tokens in the input sequence, the perplexity should take into account any other tokens in the input sequence to evaluate the likelihood of the target token.}) and is calculated by the average likelihood of a sequence. It is a common metric to evaluate how ``natural'' the text is from the LM perspective.
The perplexity of a sequence of tokens $X = (x_1, x_2, \dots, x_t)$ is defined by:
\begin{equation}
    PPL(X) = \exp \left\{- \frac{1}{t}\sum_i^t \log p_\theta(x_i | x_{\setminus i}) \right\},
\end{equation}
where $p_\theta(x_i | x_{\setminus i})$ denotes the probability of an LM $\theta$ (e.g., BERT) predicting a target token $x_i$ given the context in $X$ with $x_i$ masked out.
\hl{With the same LM, the lower perplexity score for a sentence indicates it is easier for the LM to generate the sentence.}

%% Explain how we evaluate the score
We use the perplexity to score column types for each column value (e.g., Judy Morris) by filling each column type name in the template. Then, we can evaluate if the ground truth label (i.e., ``director'' in this case) has the best (i.e., lowest) perplexity among all candidates. For the analysis, we use the vanilla BERT ({\tt bert-base-uncased}) model, which is the same base model used for \name{} in the experiments. 
We use the average rank and the normalized PPL (= PPL / Avg. PPL, where Avg. PPL denotes the average perplexity of all column types for evaluation.) 
Since perplexity values for sequences with different lengths are not directly comparable, we selected column types that are tokenized into a single token by the BERT tokenizer\footnote{Technically, column type names in the WikiTable contain hierarchical information, which is represented by URI. We used the leaf node as the column type name.}. As a result, 80 (out of 255) and 75 (out of 78) column types were selected for the WikiTable and VizNet datasets for the analysis, respectively.

\begin{table*}[t]
\caption{Language model probing results on the WikiTable dataset (Left: column type prediction. Right: Column relation prediction.) The average rank becomes $1$ if the language model always judges the column type (column relation) as the most ``natural'' choice among 80 (34) candidates for the target column value (the target column value pair.) We consider the language model has more {\em prior knowledge} about the column types (column relations) in Top-5 than those in Bottom-5.
}\label{tab:lm_probing_turl}
\begin{subfigure}[b]{0.49\textwidth}
\small
    \centering
    \begin{tabular}{c|c|cc}\hline
& Column type & Avg. rank ($\downarrow$) & PPL / Avg.PPL ($\downarrow$) \\\hline\hline
\multirow{5}{*}{\rotatebox{90}{Top-5}}& 
government.election  &   6.74 &  0.787 \\
& geography.river      &   9.25 & 0.788 \\
& religion.religion    &  10.10 & 0.799 \\
& book.author          &  12.72 & 0.810 \\
& education.university &  15.62 & 0.829 \\\hline
\multirow{5}{*}{\rotatebox{90}{Bottom-5}}& 
royalty.monarch         &  58.24 & 1.147 \\
& astronomy.constellation &  67.47 & 1.170 \\
& law.invention           &  61.60 & 1.181 \\
& biology.organism        &  71.56 & 1.205 \\
& royalty.kingdom         &  73.37 & 1.368 \\\hline
    \end{tabular}
    \footnotesize
\end{subfigure}
\begin{subfigure}[b]{0.49\textwidth}
    \centering
    \small
    %\footnotesize
    \begin{tabular}{c|c|cc}\hline
& Column relation & Avg. rank ($\downarrow$) & PPL / Avg.PPL ($\downarrow$) \\\hline\hline
\multirow{5}{*}{\rotatebox{90}{Top-5}}&
person.place\_of\_birth                       &  3.69 &   0.946 \\
& baseball\_player.position\_s                &  5.04 &   0.961 \\
& location.nearby\_airports                  &  8.66 &    0.979 \\
& mailing\_address.citytown                  &  7.24 &    0.980 \\
& film.directed\_by                              &  8.08 &  0.984 \\\hline
\multirow{5}{*}{\rotatebox{90}{Bottom-5}}& 
award.award\_nominee &  16.53 &         1.019 \\
& tv\_program.country\_of\_origin &  16.83 &         1.030 \\
& country.languages\_spoken       &  14.79 &         1.042 \\
& award\_honor.award\_winner &  21.40 &               1.047 \\
& event.entity\_involved           &  19.82 &         1.072 \\\hline
    \end{tabular}    
\end{subfigure}
\end{table*}

\begin{table}[t]
\small
    \centering
    \caption{Language model probing results on the VizNet dataset. We observe that the language model stores a certain amount of factual knowledge about column types listed in Top-5, compared to Bottom-5. The general trend is consistent with Table~\ref{tab:lm_probing_turl}.     
}
    \label{tab:lm_probing_sato}
    \begin{tabular}{c|c|cc}\hline
& Column type & Avg. rank ($\downarrow$) & PPL / Avg.PPL ($\downarrow$) \\\hline\hline
\multirow{5}{*}{\rotatebox{90}{Top-5}}& year         &   6.60 & 0.799 \\
& manufacturer &  20.19 & 0.810 \\
& day          &  14.21 & 0.819 \\
& state        &  16.88 & 0.825 \\
& language     &  17.23 & 0.840 \\\hline
\multirow{5}{*}{\rotatebox{90}{Bottom-5}}&
organisation &  61.83 & 1.146 \\
& nationality  &  65.81 & 1.218 \\
& creator      &  57.39 & 1.232 \\
& affiliation  &  63.85 & 1.239 \\
& birthPlace   &  72.30 & 1.334 \\\hline
    \end{tabular}
\end{table}

%% Relation prediction
We can use the same framework for the column relation prediction task as well. In this case, we consider a different template that has a blank field for the column relation. For example,
\begin{center}
    Derrick Henry \_\_\_\_\_ Yulee, Florida.
\end{center}
In this example, the likelihood of a sentence with ``was born in'' ({\tt place\_of\_birth}) should be judged higher than that with ``was died in'' ({\tt place\_of\_death}), which requires factual knowledge about the person. Since the column relation types are not written in plain language, we manually converted column relation type names so that they better fit in the template ({\tt entity 1}, {\tt relation}, {\tt entity 2}). Examples of converted column relation names are ({\tt place\_of\_birth}, ``was born in''), 
({\tt directed\_by}, ``is directed by''). We filtered 34 (out of 121) column relation types to make sure the converted relation type names have the exact same number of tokens.

%% Training a model from scratch
As a more direct way to test the hypothesis, we also evaluated a variant of \name{} that randomly initialized model parameters instead of using the pre-trained parameters of BERT. In this way, we can test the performance when the model with the identical architecture is trained from scratch only using training data of the target task. The model did not show meaningful performance (i.e., \hl{approximately zero F1 value}.)
We consider this is mainly because the model is too large to be trained on only the training data (i.e., without pre-training.) Thus, we decided to use the ``language model probing'' method to test the hypothesis.

\smallskip
\noindent
{\bf Results.} Table \ref{tab:lm_probing_turl} shows the results on the WikiTable dataset. We observe that some column types (e.g., goverment.election, geography.river) show lower average rank and PPL / Avg. PPL (i.e., the BERT model knows about the facts), whereas some column types (e.g., biology.organism, royalty.kingdom) show poor performance on the language probing analysis.
For example, the ``government.election'' column type is ranked at 6.74 on average and shows a smaller PPL than the average PPL. That means values in the columns that have the ``government.election'' ground-truth labels are considered ``more natural'' to appear with the term ``election''\footnote{Again, we used the leaf node of each column type as the term for the template.} than other column type names by the pre-trained LM (i.e., BERT.)
As we used 80 column types for the analysis, the ``royalty.kingdom'' column type is almost always ranked at the bottom by the LM. The poor performance could be attributed to the lower frequency of the term ``kingdom'' than other terms in the pre-training corpus.

%% Relation
For the column relations on the WikiTable dataset, the results in Table~\ref{tab:lm_probing_turl} (Right) indicate that the LM knows about factual knowledge of persons as the probing performance for relations such as {\tt place\_of\_birth} and {\tt position} is higher.
Compared to the probing results for the column types, the results show less significant differences between top-5 and bottom column relations. This is mainly because the template has three blank fields for two entities and one relation, which has a higher chance to create an unnatural sentence for the LM than that for column types.

%% Column type
The probing analysis on the VizNet dataset shows the same trend as in the WikiTable dataset. In Figure~\ref{fig:class_f1_sato}, we confirm that \name{} has better performance than Sato for all the top-5 column types. Meanwhile, {\tt birthPlace} and {\tt nationality}, which are in the bottom-5 column types for the language model probing analysis, are among the few column types where \name{} underperforms Sato. The results support that \name{} may not benefit from relevant factual knowledge stored in the pre-trained LM for the column type.

Note that the BERT model used for the analysis is {\em not fine-tuned} on the WikiTable/VizNet dataset, but the vanilla BERT model. Thus, the language model probing analysis shows the inherent ability of the BERT model, and we confirm that the pre-trained LM does store factual knowledge that is useful for the column annotation problem. 
This especially explains the significant improvements over the previous SoTA method that does not use pre-trained LM (i.e., Sato), as shown in Figure~\ref{fig:class_f1_sato}.

\section{Limitations and Future Work}\label{sec:discussion}
We have discussed why \name{} performs well for the column annotation problem through the series of analysis. In this section, we summarize the limitations of \name{} and our findings in the paper to discuss open questions for future work.

%% Table value only
\noindent
{\bf Table values only vs. with meta information.}
First, \name{} takes table value only. In most cases, we believe this assumption makes the framework more flexible to be practical.
For example, spreadsheets and data frames, which are common data format choices for data analysis, do not have table captions and often lack meaningful table headers.
Nevertheless, we acknowledge that, in some cases, meta information plays an important role to complementing table values to compose the table semantics.
As recent work~\cite{Deng:2020:TURL,Wang:2021:TCN} has shown the effectiveness of meta information for the table tasks, understanding when meta information becomes essential for the task is still an open question.

%% Assumption: a table contains multiple tables
\noindent
{\bf Single-table model vs. multi-table model.}
Second, \name{} assumes the input table to be {\em self-contained.} That means columns that are necessary to compose table context should be stored in the same table.
Web Tables generally follow the assumption, and \name{} shows the strong performance on the WikiTable and VizNet datasets.
However, \name{} was not tested on relational tables, where chunks of information can be split into multiple tables after database normalization.
In such a scenario, we need to consider inter-table relations and information to incorporate key signals outside the target table.
In fact, contemporaneous work~\cite{Wang:2021:TCN} has developed a framework that incorporates signals from external tables for better column annotation performance.
Thus, we consider joint modeling of multiple tables should be a future direction.

%% Dirty data, although it is known that pre-trained models are known to be robust for dirty data. 
\noindent
{\bf Clean data vs. dirty data.}
Third, our framework assumes that table values are ``correct and clean'', which may not always be true in real-world settings.
The input table value should be of the high-quality, especially when we limit the max input token size for better efficiency.
Recent studies that applied pre-trained LMs to tables~\cite{Li:2020:Ditto,Li:2021:JDIQ:DeepEntityMatching} have shown that the pre-trained LM-based approach achieves robust improvements even on ``dirty'' datasets, where some table values are missing or misplaced.
Following the error detection/correction research~\cite{Mahdavi:2019:Raha,Mahdavi:2020:Baran}, which has been studied independently, implementing functionality that alleviates the influence from the incorrect table values is part of the future work.

%% More tasks, and task similarity.
\noindent
{\bf Multi-task learning with more tasks.}
Lastly and most importantly, there are many open questions in applying multi-task learning to data management tasks. Although we have shown that multi-task learning is useful for the column annotation task, it is not yet very clear what types of relevant tasks are useful for the target task.
A line of work in Machine Learning has studies on the task similarity and the transferability of the model~\cite{Taskonomy2018}.
Therefore, it is also important for us to understand the relationship between task similarity and benefits of the multi-task learning framework.
We acknowledge that our study is still preliminary with respect to this point. However, we believe that our study established the first step in this research topic toward a wider scope of multi-task learning applicability to other data management tasks.

\fi

\end{document}